\documentclass[manuscript]{aastex}
\usepackage{times}
\usepackage{color}

\newcommand{\Suk}[1]{{#1}}
\newcommand{\SC}[1]{{#1}}
\newcommand{\RP}[1]{{#1}}
\usepackage{graphicx}
\usepackage[flushleft]{threeparttable}
\usepackage{soul}   
\usepackage{enumerate}
\usepackage{footnote}

\tabletypesize{\scriptsize}

\shorttitle{The Extended Virgo Cluster Catalog}
\shortauthors{Kim et al.}

\begin{document}
\title{The Extended Virgo Cluster Catalog}
\author{Suk Kim\altaffilmark{1}, Soo-Chang Rey\altaffilmark{1}$^{,}$\altaffilmark{5}, Helmut Jerjen\altaffilmark{2},Thorsten Lisker\altaffilmark{3}, Eon-Chang Sung\altaffilmark{4}, Youngdae Lee\altaffilmark{1}, Jiwon Chung\altaffilmark{1}, Mina Pak\altaffilmark{1}, Wonhyeong Yi\altaffilmark{1}, and Woong Lee\altaffilmark{1}}
\altaffiltext{1}{Department of Astronomy and Space Science, Chungnam National University, 99 Daehak-ro, Daejeon 305-764, Korea}
\altaffiltext{2}{Research School of Astronomy and Astrophysics, The Australian National University, Cotter Road, Weston, ACT, 2611, Australia}
\altaffiltext{3}{Astronomisches Rechen-Institut, Zentrum f\"ur Astronomie der Universit\"at Heidelberg (ZAH), M\"onchhofstra\ss e 12-14, D-69120 Heidelberg, Germany}
\altaffiltext{4}{Korea Astronomy \& Space Science institute, 776 Daedeokdae-ro, Daejeon 305-348, Korea}
\altaffiltext{5}{Author to whom any correspondence should be addressed}

\begin{abstract}
We present a new catalog of galaxies in the wider region of the Virgo cluster, based on the Sloan Digital Sky Survey (SDSS) Data Release 7. The Extended Virgo Cluster Catalog (EVCC) covers an area of 725\,deg$^{2}$ or 60.1\,Mpc$^2$. It is 5.2 times larger than the footprint of the classical Virgo Cluster Catalog (VCC) and reaches out to 3.5 times the virial radius of the Virgo cluster. We selected 1324 spectroscopically targeted galaxies with radial velocities less than 3000\,km\,s$^{-1}$.  In addition, 265 galaxies that have been missed in the SDSS spectroscopic survey but have available redshifts in the NASA Extragalactic Database are also included. Our selection process secured a total of 1589 galaxies of which 676 galaxies are not included in the VCC. The certain and possible cluster members are defined by means of redshift comparison with a cluster infall model. We employed two independent and complementary galaxy classification
schemes: the traditional morphological classification based on the visual inspection of optical images and a characterization of galaxies from their spectroscopic features. SDSS $u$, $g$, $r$, $i$, and $z$ passband photometry of all EVCC galaxies  was performed using Source Extractor. We compare the EVCC galaxies with the VCC in terms of morphology, spatial distribution, and luminosity function. The EVCC defines a comprehensive galaxy sample covering a wider range in galaxy density that is significantly different from the inner region of the Virgo cluster. It will be the foundation for forthcoming galaxy evolution studies in the extended Virgo cluster region, complementing ongoing and planned Virgo cluster surveys at various wavelengths.
\end{abstract}

\section{INTRODUCTION}

 The current theory of structure formation in the standard cold dark matter universe predicts that the build-up of galaxy clusters is characterized by the accretion of individual galaxies and groups. These small galaxy aggregates preferentially travel from the cosmic web along filamentary substructures to the densest regions \citep{van93,Bon96,Spr05}. Nearby galaxy clusters ($cz< 10000$\,km\,s$^{-1}$) and their surrounding regions represent the current endpoint of this evolution and occupy the apex of the process. 

However, despite the importance of these extended cluster regions most observational studies of nearby galaxy clusters have been concentrating on the densest parts, the cores or central regions that are already in dynamical equilibrium. There are comparatively few examinations of the extended infall regions of clusters due to the lack of wide-field CCD cameras and spectroscopic information. These intermediate density environments contain the structural bridges that lead to the cluster cores. It is expected that because of the longer dynamical time scale they still contain important information about the different stages of the cluster formation and the physical processes that change the morphologies of galaxies. In this regard, it is intriguing to study the galaxy populations that we encounter in the extended areas of nearby clusters.

While high quality data for distant clusters is continuously being gathered from the Hubble Space Telescope (HST) and large ground-based telescopes, our present knowledge of the systematic properties of galaxies in nearby clusters remains surprisingly limited. Photography-based surveys of clusters in Virgo \citep{Bin85}, Fornax \citep{Fer89}, Centaurus \citep{Jer97_2}, and Coma \citep{God83} are still cited as some of the main references. 
The significance of these surveys can be understood given the fact that large format, wide-field CCD imagers that enable the mapping of a significant number of nearby clusters became available only recently (e.g., \citealt{Fer12} for Virgo; \citeauthor{Pak14} in prep. for Ursa Major). With the availability of imaging and spectroscopic data from megasurveys such as the Sloan Digital Sky Survey (SDSS), the situation of studies for nearby clusters in the northern hemisphere has been revolutionized. The SDSS data has the advantage of uniform coverage over large areas of sky with photometric and spectroscopic depths that are suitable to study detailed properties of nearby cluster galaxy populations.

The Virgo cluster is the nearest rich cluster from the Milky Way at a distance of 16.5\,Mpc \citep{Jer04,Mei07}. The Virgo Cluster Catalog \citep[VCC,][]{Bin85} has been extensively used and is the primary source catalog for countless studies of properties of cluster galaxies. It contains 2096 galaxies that are distributed within an area of approximately one cluster virial radius \citep[R$_{vir}$ = 6\,deg or 1.72\,Mpc;][]{Hof80}. As the cluster environment already starts affecting the evolution of galaxies located at several virial radii from the center and the Virgo cluster is considered to be a dynamically young cluster \citep{Arn04,Agu05}, investigations of the more extended regions of the Virgo cluster should still contain valuable information about the mass assembly history of the cluster, the physical processes that govern the morphological transformation of individual galaxies, and the connection of the galaxy distribution with the large scale filamentary structures around the Virgo cluster \citep{Tul82}. 

In line with the above situation, we present a new catalog of galaxies in the surrounding region of the Virgo cluster, what we call the Extended Virgo Cluster Catalog (EVCC), using publicly available data from the Sloan Digital Sky Survey (SDSS). The SDSS provides multi-band optical images that are a critical asset for the investigation of galaxy properties. These multi-color images also allow good quality of the morphological classification and in some cases the improvement of the morphological classification of VCC galaxies, which was based on large-scale, blue sensitive photographic plates with an angular resolution of 10.8\,arcsec mm$^{-1}$. While the membership determination of galaxies in the VCC relied primarily on the visual inspection of the single band images and galaxy surface brightness, the SDSS can also take advantage of color information and the spectroscopic data for refined classification of galaxies in the Virgo cluster region.

We organize this paper as follows. Section 2 describes the definition of the EVCC region and the selection of galaxies. In section 3, we introduce two schemes of galaxy classification based on the SDSS imaging and spectroscopic data. Galaxy photometry is described in section 4 and the final EVCC catalog is presented in section 5. In section 6, we present a comparison between the EVCC and VCC. Finally, we summarize our results in {section 7}.

\section{SELECTION OF GALAXIES}
    \subsection{Data}
    The construction of the EVCC is based on the SDSS Data Release 7 (DR7) \citep{Aba09}. The SDSS DR7 provides reduced and calibrated images taken in the $u$, $g$, $r$, $i$, and $z$ bands with an effective exposure time of 54s in each band \citep[see also][]{Sto02}. The pixel scale of 0.396\,arcsec and the average seeing of 1.4\,arcsec correspond to a physical size of 32\,pc and 112\,pc, respectively, at a Virgo cluster distance of 16.5\,Mpc \citep[i.e., a distance modulus $m - M = 31.1$;][]{Jer04,Mei07}. The SDSS spectroscopic survey covers nominally all galaxies brighter than $r\leq$ 17.77 and $r$-band Petrosian half-light surface brightnesses $\mu_{50} \leq 24.5$\,mag\,arcsec$^{-2}$ \citep{Str02}. It provides fiber spectra with a wavelength coverage between 3800 - 9200\AA\, at a resolution of $R=2000$. The fiber has an angular diameter of 3\,arcsec that corresponds to 240\,pc at a distance of Virgo, providing spectral information for the very central region of the galaxies. We took advantage of the SDSS spectroscopic data for the determination of galaxy membership and classification of spectroscopic morphology (see Sec. 2.4 and Sec. 3.2). The SDSS imaging data allow detailed morphological classification and accurate photometry of galaxies (see Sec. 3.1 and Sec. 4 for the details).

	\subsection{Region of the EVCC}
\begin{figure}
    		\begin{center}
    		\epsscale{1}
    		\plotone{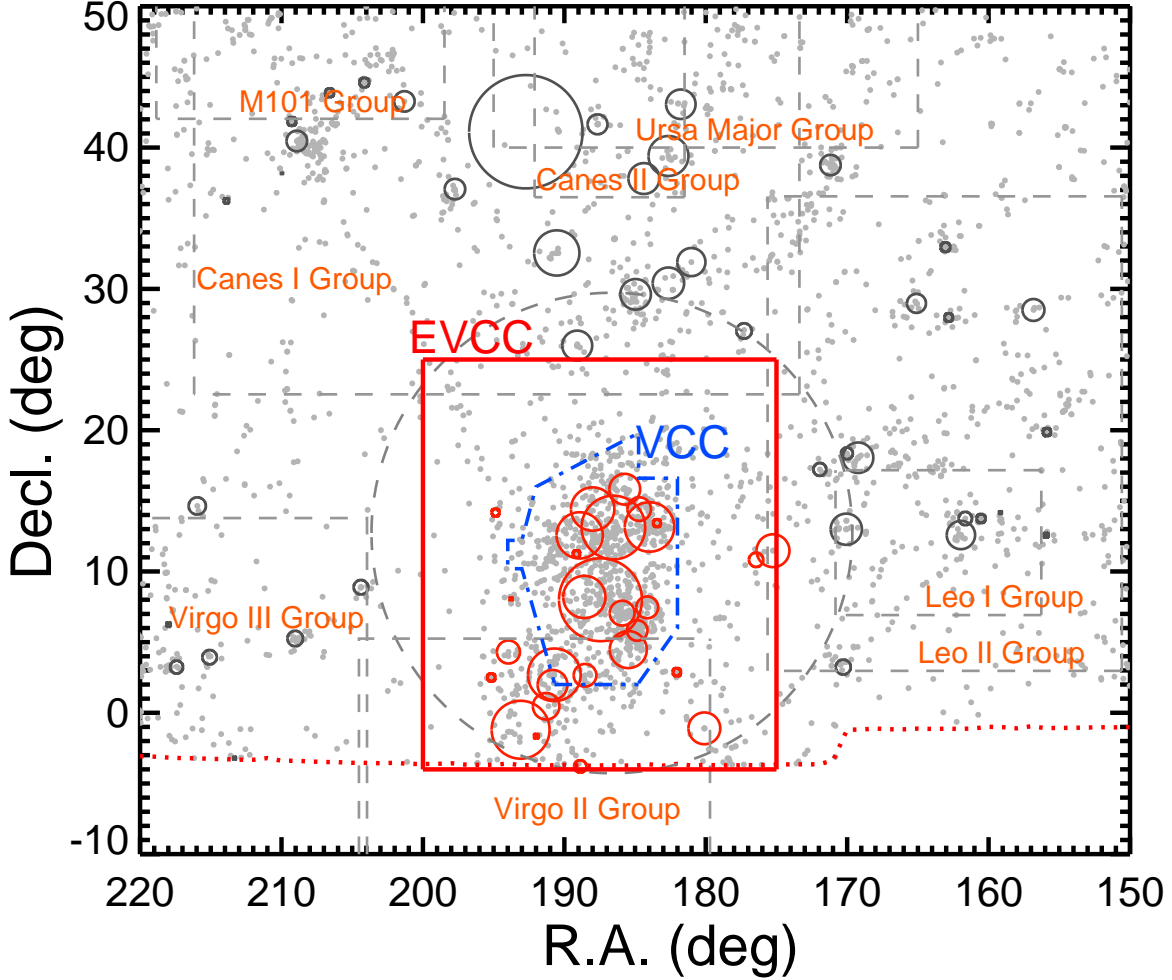}
    		\caption{Projected spatial distribution of galaxies in the direction of the Virgo cluster and its surrounding cosmic web. Grey dots are all galaxies from the SDSS DR7 spectroscopic data with radial velocities less than 3000\,km\,s$^{-1}$. The large red rectangle denotes the region of the EVCC, and the region of the VCC is presented by the blue dot-dashed contour. Grey dashed line rectangles are outlines of other galaxy groups surrounding the Virgo cluster. Red and grey circles are galaxy groups which are located inside and outside of the EVCC region, respectively, from the galaxy group catalog of \citet{Mak11}. The circle size is proportional to the mean harmonic radius of the group from \citet{Mak11}. For reference, we also present the lower limit of the zero-velocity surface ($\sim$17 deg radius, large dashed circle) estimated by \citet{Kar10}. \RP{The red dotted line indicates the coverage limit of the SDSS observations.}
}  
         \label{DATA_EVCC_PSD}
    		\end{center}
    		\end{figure}

	The SDSS spectroscopic data was searched in the window of 175\,deg $<$ R.A.\,(J2000) $<$ 200\,deg and -4\,deg $<$ Decl.\,(J2000) $<$ 25\,deg, which defines the nominal area of the EVCC (large red rectangle in Figure~\ref{DATA_EVCC_PSD}). This region extends to  the lower limit of the zero-velocity surface of the Virgo cluster, about 17\,deg radius from the cluster center, 
estimated by \citet{Kar10}, which separates the galaxies in a given galaxy cluster against the global cosmic expansion. In contrast, the VCC region covers a smaller area within a radius of $\sim$10 deg (blue dot-dashed contour in Fig.~\ref{DATA_EVCC_PSD}). The EVCC area (725 deg${^2}$) is 5.2 times larger than the footprint of the VCC (140 deg${^2}$), reaching up to 3.5 times the virial radius of the Virgo cluster. 
	
	The EVCC region generally avoids overlap with other distinct galaxy groups in the vicinity \citep[grey dashed line rectangles in Fig.~\ref{DATA_EVCC_PSD},][]{Fou92,Gar93,Giu00}.
	From the nearby galaxy group catalog of \citet{Mak11}, the EVCC region includes a complex of subgroups around the main body of the Virgo cluster (red circles in Fig.~\ref{DATA_EVCC_PSD}) and the neighboring galaxy groups are also located outside of the EVCC region (grey circles in Fig.~\ref{DATA_EVCC_PSD}). The southern extension of the Virgo cluster at Decl. $<$ 5\,deg, which is a part of the Local Supercluster \citep{deV73,Tul82,San85_3,Hof95}, is well covered by the EVCC, while the VCC region included only the most northern tip of this structure. However, the EVCC does not cover the area below Decl. = -4\,deg where the Virgo II Group \citep{deV61,deV73} is located, owing to the coverage limit of the SDSS observations (see red dotted line in Fig.~\ref{DATA_EVCC_PSD}).

	\subsection{Selection of Sample Galaxies}
\begin{figure}
		\begin{center}
	  	\epsscale{1}
	  	\plotone{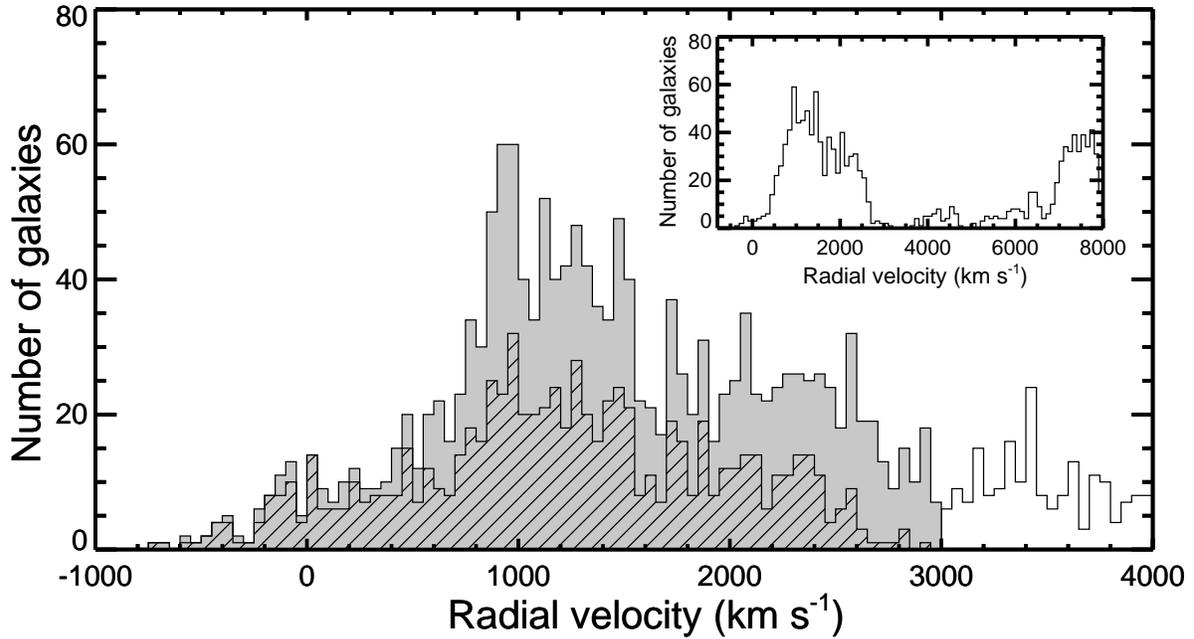}	
	  	\caption{
       \RP {Radial velocity distribution of all galaxies in the region of the EVCC (grey and open histograms) from the SDSS spectroscopic data. The velocity distribution of galaxies in the EVCC region  shows a dip around 3000\,km\,s$^{-1}$. The hatched histogram is for galaxies listed in the VCC catalog. The inset shows a histogram of galaxies from the SDSS spectroscopic data located in the VCC region. A gap in this distribution around 3000\,km\,s$^{-1}$ is clearly seen and used to separate cluster members from background. The selected galaxies included in the EVCC are those with radial velocities less than 3000\,km\,s$^{-1}$ (grey histogram).}
	  	}       
	  	\label{DATA_EVCC_VELD}
		\end{center}
		\end{figure}

   We initially focus on galaxies with radial velocities that are available in the SDSS spectroscopic data since the velocity is a suitable information for cluster membership of galaxies in a galaxy cluster without significant fore- and background. \RP{In Figure~\ref{DATA_EVCC_VELD}, we present the velocity distribution of galaxies in the region of the EVCC (grey and open histograms).} We selected 1324 galaxies with radial velocities less than 3000\,km\,s$^{-1}$ in the EVCC region \RP{(i.e., grey histogram).} The value of 3000\,km\,s$^{-1}$ was adopted as an upper limit in the velocity distribution of the Virgo galaxies since a well-defined gap around 3000\,km\,s$^{-1}$ has been acknowledged by \citet{Bin93} \RP{(see inset of Fig.~\ref{DATA_EVCC_VELD}), while galaxies in the EVCC region exhibit a dip in their radial velocity distribution} owing to the contribution of galaxies outside of the VCC region.
	
	While SDSS spectroscopic observations generally cover galaxies down to an apparent magnitude of  $r$ = 17.77, bright, bulge-dominated galaxies with $r$ $<$ 14.5 were rejected from the SDSS spectroscopic target list as they cause saturation problems \citep{Bla05_1, Bla05_2}. Another source of incompleteness are fiber conflicts, where neighboring fibers could not be placed closer than 55\,arcsec \citep{McI08}. To complete the EVCC catalog with galaxies for which radial velocities were not measured in the SDSS, we extracted independent spectroscopic data from the NASA/IPAC Extragalactic Database (NED) for 265 galaxies that are located in the EVCC region. These galaxies cover a wide range in magnitude, 8 $<$ $r$ $<$ 19. \RP{As a result, the EVCC contains a total of 1589 galaxies of which 676 galaxies are included only in the EVCC (see also Figure~\ref{onlyEVCCpsd} and Sec. 6.4).}

  To test the consistency between SDSS and NED, we examined the radial velocities of 498 EVCC galaxies which are available in the SDSS and NED. The mean difference (2.6\,km\,s$^{-1}$) and standard deviation (105\,km\,s$^{-1}$) of radial velocity turn out negligible. Furthermore, we found that there is no significant deviation of the radial velocity depending on the magnitude range. Specifically, faint galaxies ($r$ $>$ 16) also show similar residuals (i.e., 10 $\pm$ 103\,km\,s$^{-1}$). Consequently, we conclude that radial velocity information of our sample is not significantly hampered by difference of compiled sources. 
    
  The spectra of most SDSS galaxies are obtained from fibers positioned on or close to the galactic center, but some galaxies in the SDSS have their spectrum taken at an \Suk{off-center} position. Some galaxies have even multiple spectra observed at various \Suk{off-center} positions. In the latter case, we select the fiber spectrum obtained at the location closest to the center to minimize uncertainties of radial velocities due to rotation or peculiar motion within a galaxy. About 10$\%$ of the SDSS data included in the EVCC have \Suk{off-center} spectra.
    
    We also examined the reliability of radial velocities obtained from \SC{off-center} spectra. We select galaxies whose spectra from the SDSS are off-centered by more than 5\,arcsec from the photometric galaxy center and compare their radial velocities with independent measurements in the NED where spectra are observed on their center. We found that the radial velocities of these galaxies show no significant difference with those from the NED (i.e., 12 $\pm$ 98\,km\,s$^{-1}$), indicating no appreciable systematic bias with respect to radial velocities from \SC{on-center} spectra. 

\begin{figure}
		\epsscale{1}
		\begin{center}
		\plotone{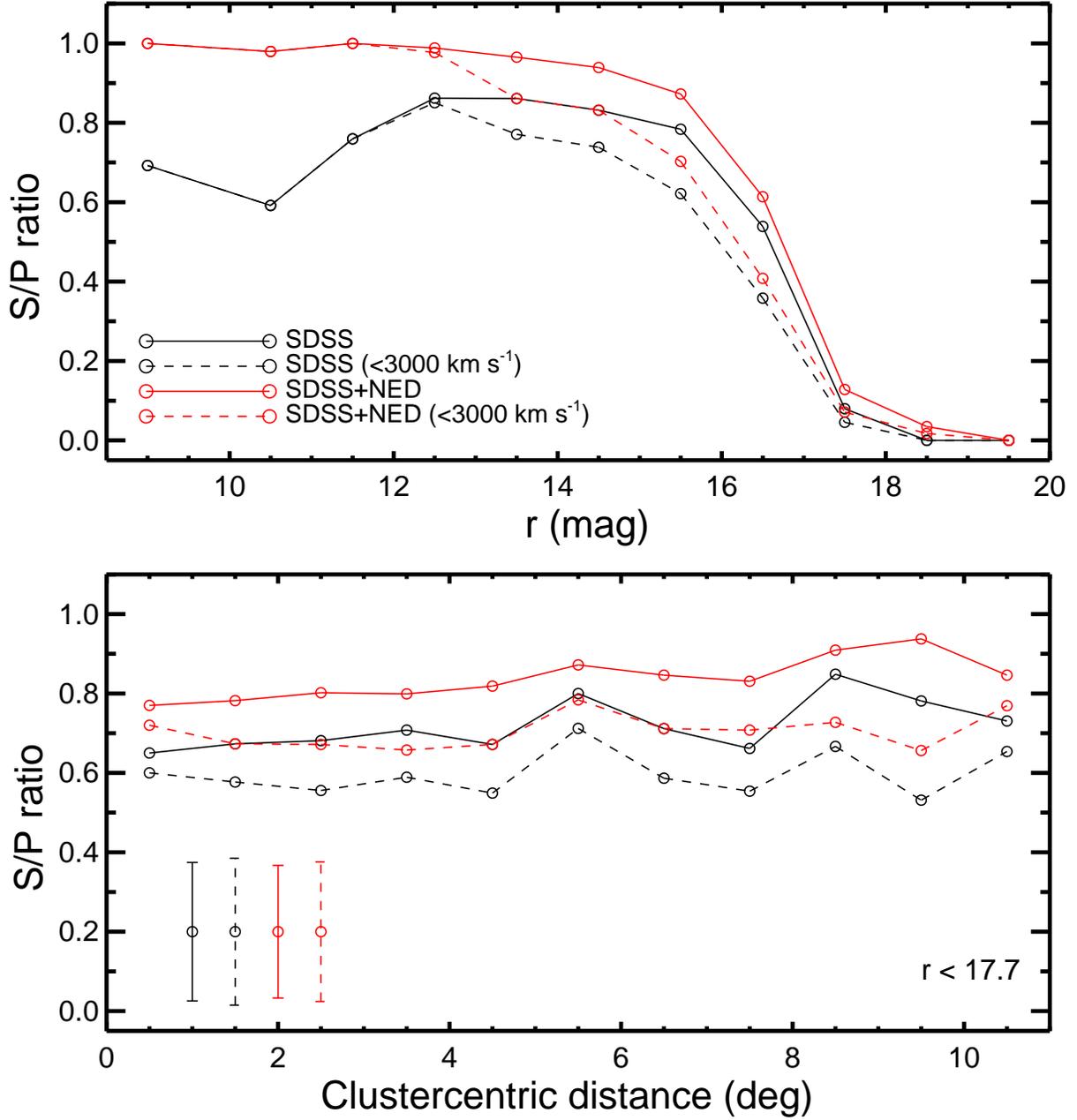}
		\caption{ (Top) The S/P ratio (the fraction of certain and possible VCC member galaxies with available spectroscopic data) against the r-band magnitude. (Bottom) The S/P ratio of galaxies brighter than  r = 17.7 with respect to the clustercentric distance from the Virgo cluster center. The red and black lines indicate S/P ratios obtained from the SDSS data with and without the NED data, respectively. Dashed and solid lines indicate galaxies with $cz$ $<$ 3000\,km\,s$^{-1}$ and with all redshift data, respectively. \RP{Error bars in the bottom panel denote the mean values of Poisson error in  different S/P ratios.}
		 }
		\label{fraction}
		\end{center}
		\end{figure}

    In Figure~\ref{fraction}, we estimate the number ratio of VCC certain and possible member galaxies with available radial velocities from \Suk{SDSS and NED} to those with the SDSS photometric data (hereafter S/P ratio) as a function of magnitude (top panel) and clustercentric distance (bottom panel). The S/P ratio is obtained for the SDSS data with (red solid line) and without (black solid line) the NED data. The S/P ratio of galaxies with $cz$ $<$ 3000\,km\,s$^{-1}$ (i.e., S/P$_{3000}$ ratio, dashed lines) is also considered. The overall feature of the S/P$_{3000}$ ratio versus the magnitude is that it drops to the 50 percent completeness level around r $\sim$16.5 (M$_{r}=-14.6$). Galaxies fainter than $r=18$ (M$_{r}=-13.1$) remain largely undetected by the SDSS and NED data. In the case of S/P ratio just for the SDSS data, a prominent paucity of spectroscopic data for bright galaxies with $r$ $<$ \RP{13} is seen. This highlights the saturation problem of SDSS spectroscopic observations mentioned by \citet{Bla05_1, Bla05_2}. 
    
    However, by adding the NED data, the S/P ratio increases to $\sim$100$\%$ for about $r<$ 14. At a given magnitude, the S/P$_{3000}$ ratio is lower than the S/P ratio in the magnitude range of 13 $< r <$ 18. This indicates that some fainter galaxies in the morphology-based VCC are absent in the EVCC because their redshifts are larger than 3000\,km\,s$^{-1}$ (see also Sec. 6.4).
    
    In addition, we examine the S/P ratio for galaxies brighter than the magnitude limit of SDSS spectroscopic observations (i.e., $r =$ 17.7) \Suk{with respect to the clustercentric distance from \SC{the} Virgo} (see bottom panel of Fig.~\ref{fraction}). The S/P ratio shows no dependence on the location within the cluster, \RP{considering Poisson error ($\sim\pm$0.18) in the S/P ratio (see error bars in the bottom panel of Fig.~\ref{fraction}).} This suggests that the average angular separation \Suk{between galaxies at the distance of the Virgo cluster} is large enough to avoid the problem of fiber collisions in the SDSS observations \citep[i.e., 55\,arcsec;][]{Str02}.    
		
	\subsection{Membership}
\RP{Cluster membership determination of galaxies in the VCC was essentially based on optical morphology applying some criteria (see \citealt{San84,Bin85} for the details). Cluster member early-type dwarfs and irregular galaxies are well distinguished from the background galaxies, since they follow a well established correlation between magnitude and surface brightness \citep{Bin91,Bos08}. Thus, at a given magnitude, these cluster members exhibit lower surface brightness than the majority of background counterparts. \Suk{In the case of spiral galaxies, luminosity classes combined with apparent galaxy size were used to distinguish members from} background objects \citep{van60}. Additionally, \Suk{resolution into star-forming knots} provided a complementary criterion for late-type galaxies; i.e., member galaxies exhibit well resolved HII regions and stellar associations compared to the background counterparts.} \citet{Bin93} subsequently demonstrated with newly available radial \Suk{velocity data that morphology is a powerful method} to separate cluster members and background galaxies with a 95\% success rate. These authors noted that in the case of the Virgo cluster membership status can be best determined with radial velocity data because of a distinct gap in the velocity distribution behind the Virgo cluster at 3000\,km\,s$^{-1}$.

		\begin{figure}
		\epsscale{1}
		\plotone{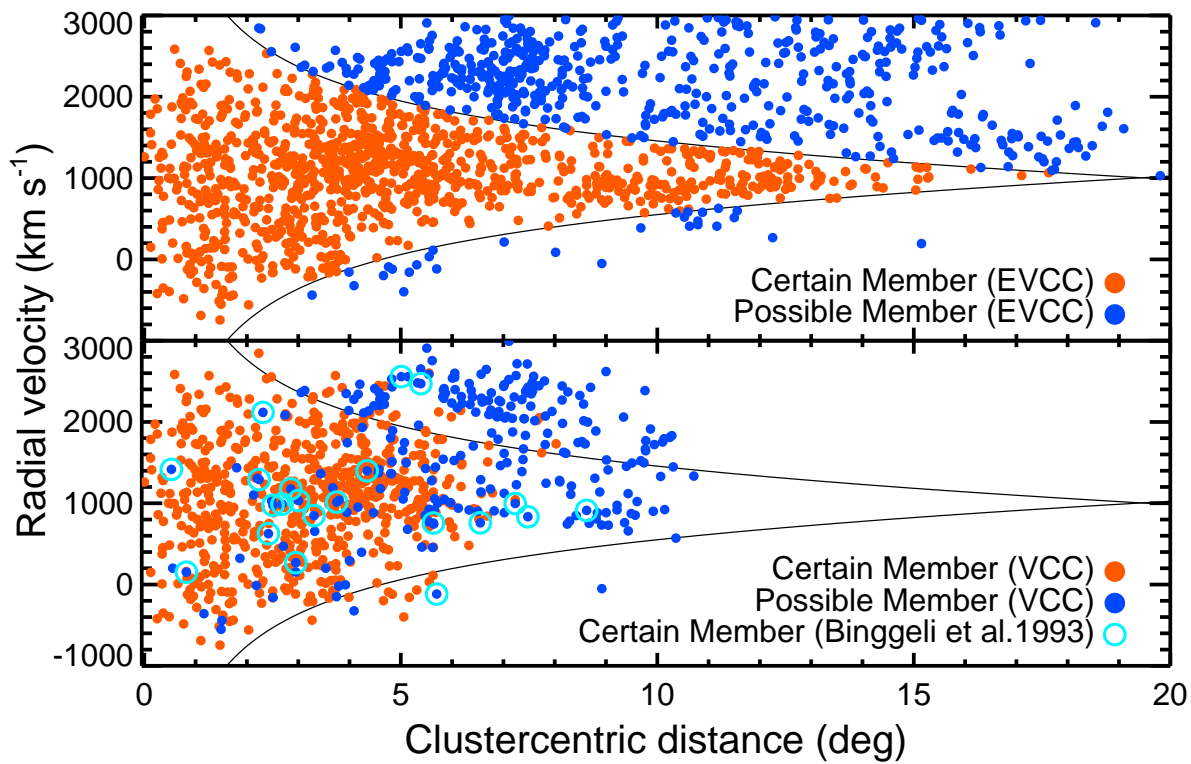}
		\caption{
		(Top) Radial velocities of EVCC galaxies against projected radius from the cluster center. The solid lines indicate the model caustic curves using the infall model of \citet{Pra94}. Red and blue dots are member and possible member galaxies defined as galaxies located inside and outside of the model lines, respectively. (Bottom) Radial velocities of VCC galaxies against projected radius from the cluster center. Membership is taken from the VCC and radial velocity information is from the SDSS or NED. Cyan open circles are possible member galaxies in the VCC which were re-classified  as certain members based on the radial velocity information from \citet{Bin93}.
		}
    \label{Sel_member}
	\end{figure}

    We divide the EVCC galaxies into two subsamples using a spherical symmetric infall \Suk{model that defines the limits of the infall velocity} as a function of distance from the cluster center \citep{Pra94}.
    \Suk{\RP{Under the assumption that galaxy clusters are small density perturbations in the early universe, a spherical infall model predicts the galaxy motions based on the spherical accretion onto a point mass in an uniform and expanding universe. The density perturbations eventually have broken away from the general Hubble expansion as a result of the dominating gravitational force out to a maximum distance (i.e., turnaround radius), allowing material to move inwards. The infall motion produces a caustic curve in a galaxy cluster which consists of all galaxies whose infall motion is larger than the local Hubble flow. The spherical infall model can be calculated from the virial radius and the line-of-sight velocity dispersion of a cluster \citep{Pra94,Abd11}.}}
\RP{Consequently,} a caustic curve for a model of the Virgo cluster is obtained using $\Omega$${_0}$ = 0.3 \citep{Jin95}, a velocity dispersion ${\sigma}$${_v}$ = 800\,km\,s$^{-1}$ \citep{Bin93}, and a virial radius of 1.8\,Mpc \citep{Hof80}. 

In Figure~\ref{Sel_member} (top panel), we plot the radial velocities of \Suk{the} EVCC galaxies against Virgocentric distance, measured from M87, along with the model caustic curves. We define certain member (red dots) and possible member galaxies (blue dots) as those located inside and outside of the model lines, respectively. For comparison, we also present VCC galaxies in Fig.~\ref{Sel_member} (bottom panel) using membership classification from the VCC. Most VCC galaxies classified as certain member (red dots) based on the morphology are inside of the model lines, confirming they are true members in terms of their radial velocity information. It is worth to note that some possible member (blue dots) galaxies classified in the VCC are also enveloped in the caustic curves showing relatively small deviation of their radial velocities to the cluster's systemic velocity. \citet{Bin93} investigated the membership of 144 galaxies in the VCC using available radial velocity information at that time. As a result, about 22\% (21 of 94) of possible member galaxies were re-classified as certain member galaxies \citep[see Table 1 of ][]{Bin93}. Note that all but three of these galaxies are located inside of the caustic curves (cyan circles in \SC{the} bottom panel of Fig.~\ref{Sel_member}).
    
    By definition, galaxies that are located outside of the caustic curves of the infall model should be considered as non-member galaxies of the Virgo cluster. Nevertheless, \SC{the} projected spatial distribution of these galaxies appear to be similar to that of possible member galaxies in the VCC (see Figure~\ref{Sel_member_field} of Sec. 6.2). Furthermore, in the VCC, substructures of the Virgo cluster, including the W, W$'$, and M clouds and southern extension region, are dominated by possible member galaxies \citep{Bin85,Bin87,Bin93,San85_3}, and some results suggest that they are connected with the M87 and M49 clusters (see \citealt{Yoo12} and references therein). Therefore, we characterize galaxies located outside of the infall model lines as possible members, following the VCC terminology.

\section{MORPHOLOGY CLASSIFICATION}  
The SDSS data allow us to investigate the relationship between photometric and spectroscopic properties of galaxies. We introduce two classification schemes using the SDSS imaging and spectroscopic data. In addition to the traditional morphological classification by visual inspection of images (``primary morphology"), we also classify galaxies based on their spectroscopic features (``secondary morphology"). The primary morphological classification follows the scheme of the morphological classification used in the VCC. The secondary morphological classification relies on the spectral energy distribution (SED) shape and presence of H${\alpha}$ emission/absorption lines. The morphological classification for all galaxies was independently performed by three members of our team (H.J., T.L., and E.C.S.).

		\subsection{Primary Morphology}
\begin{figure}[htp]
        \epsscale{1}
        \plotone{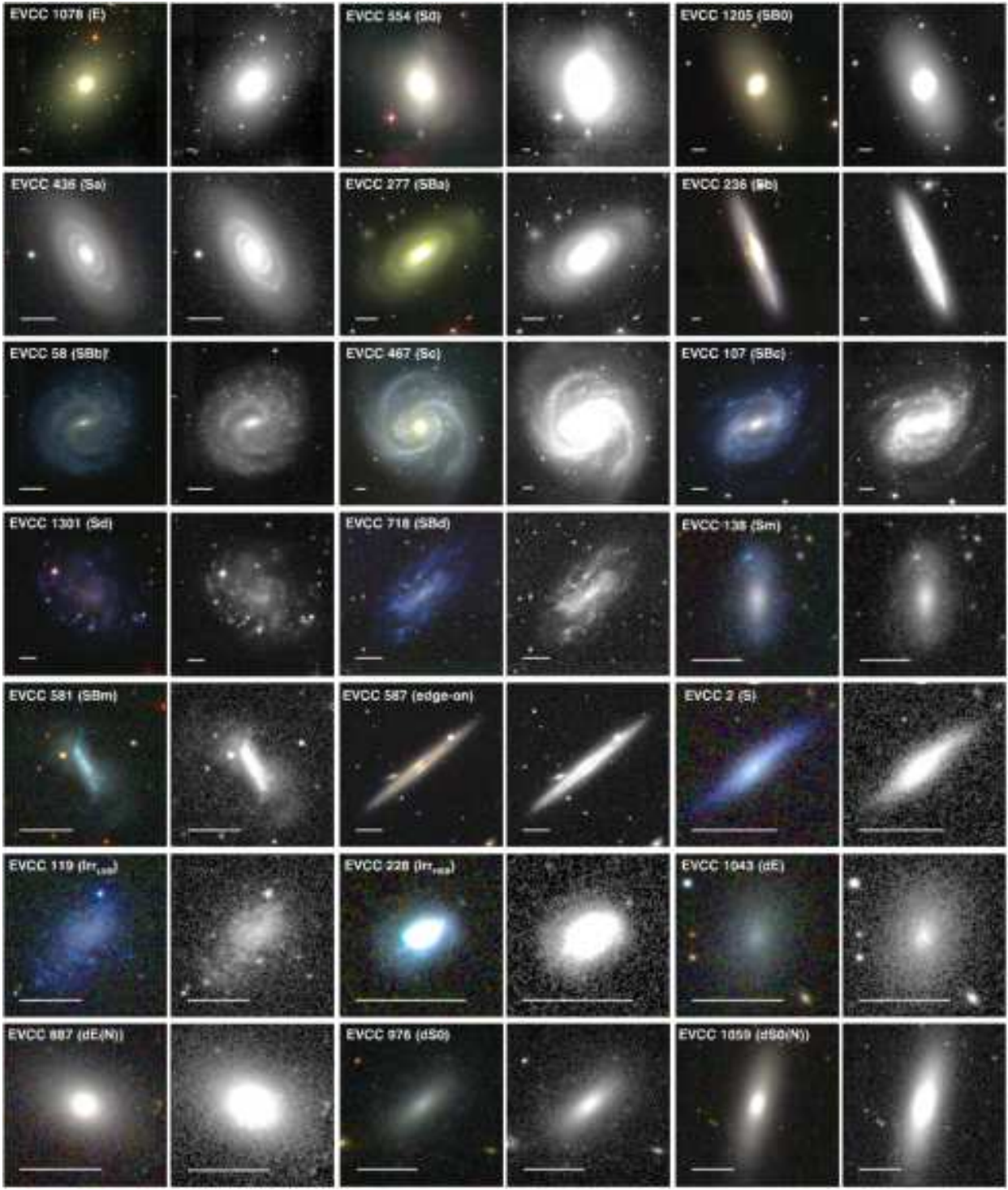}
		  \caption{
		  Examples of galaxies with various primary morphologies. For each galaxy we show the $gri$ combined color image (left) and  the $r$-band grey scale image (right). In each panel, the solid bar corresponds to 30\,arcsec (2.4\,kpc) on the sky. North is up and east is to the left. The EVCC number and the morphological type are also given in the upper left corner of the color image.
		  }
		  \label{Pri_mor}
		\end{figure}

		The morphological classification was carried out by carefully examining both monochromatic $g$, $r$, and $i$-band images and $gri$ combined color images of the SDSS data. Some of the advantages of using digital images over photographic plates is the ability to zoom in, to vary contrast levels as well as cut levels to investigate low and high surface brightness features in details. The EVCC galaxies were subdivided into 21 classes on the basis of the extended Hubble morphological classification scheme. These classes are summarized in Table~\ref{pmor} and illustrated in Figure~\ref{Pri_mor}. We briefly describe the most important features here and refer to the following main references for a comprehensive discussion and illustration of the different types: \citet{San84,San85_2,Bin91,San94}\footnote{http://ned.ipac.caltech.edu/level5/Shapley\_Ames/frames.html}.

        \begin{table}[hpt]
        \caption{Primary Morphology}
        \resizebox{17cm}{!}
        {
        \begin{tabular}{lll}
        
        \hline
        \multicolumn{1}{c}{Type} & \multicolumn{1}{c}{Digital Code} & \multicolumn{1}{c}{Subclass} \\
        \hline
        Elliptical & 100 &  \\
        Disk galaxy &	2[0,1][0-7]  & [no-bar=0/bar=1] [subtype: 0=0, a=1, b=2, c=3, d=4, m=5, S=6, edge-on=7] \\
        Irregular	 &	3[0,1]0 & [number of prominent HII regions: none to low=0, high=1] \\
        Early-type dwarf galaxy & 4[0,1][0,1]  & [subtype: dE=0, dS0=1] [no nucleus=0/nucleus=1] \\\hline
        \end{tabular}
        }
        \label{pmor}
        \end{table}

	\begin{enumerate} [-]
	   \item Elliptical galaxy (E) : Elliptical galaxies show high surface brightness spheroidal shape and a featureless envelope.    
	   \item Lenticular galaxy (S0) : The main feature that discriminates between an E and a S0 galaxy is the presence of a disk. S0 galaxies have a smooth bulge and disk component with a distinct change in surface brightness at the transition radius when seen face-on. Edge-on S0s display a lens shape.
	   \item Spiral (Sa - Sd) : The classification criteria for spiral galaxies are the relative strength of the bulge and the degree of tightness of the spiral arms defined by bright HII regions. Qualitatively, early-type spirals (e.g., Sa and Sb) have a large bulge and tightly wrapped spiral arms, while late-type spirals (e.g., Sc and Sd) have a small bulge and patchy open spiral arms.
	   \item Early-type dwarf galaxy (dE and dS0) : A low surface brightness galaxy characterized by elliptical isophotes and a smooth surface brightness profile.  Overall colors are generally red, but some faint dEs show blue colors. Among the dE class, some galaxies show a distinct bulge-disk transition and/or asymmetric features (e.g. EVCC 976 in Fig.~\ref{Pri_mor}). They are classified as dS0. If an unresolved, star-like nucleus at the center is present in a dE or a dS0, they are called dE(N) or dS0(N).
      \item Sm : Sm galaxies are termed the Magellanic type spirals and are the latest type where traces of spiral patterns are still visible. 
      \item Barred galaxy : Barred galaxies exhibit a large-scale linear bar feature crossing the nucleus.
      \item High surface brightness irregular (HSB Irr) and low surface brightness irregular (LSB Irr) : Irregular galaxies have a chaotic morphology. We use the surface brightness to separate the blue compact dwarf (BCD) type from the ``Im" galaxies similar to the luminosity class in the Revised Shapley-Ames Catalog \citep{San81}. The EVCC irregulars of highest surface brightness are called HSB Irr. Their appearance is dominated by a single, or sometimes group of bright, centrally located HII regions. We classified all irregular EVCC galaxies as if they were at the Virgo cluster distance (assuming an appropriate line-of-sight depth). However HSB galaxies are the most difficult to get right using just the redshift and angular size. It is possible that there are a few background spiral galaxies in the HSB class. The LSB Irr galaxies are irregular galaxies with low surface brightness similar to dEs but with irregular isophotes. \RP{Based on the $r$-band peak surface brightness values and half-light radius (R50) returned from \SC{SExtractor}, at all given magnitudes, HSB Irr galaxies \Suk{are} found to have systematically high central surface brightness (difference of about 1\,mag\,arcsec$^{-2}$) and small angular size (a factor of about 2 in R50) compared to the LSB Irr counterparts.}   
      \item Edge-on galaxy : Edge-on galaxies are spiral galaxies seen edge-on and thus did not allow a more detailed morphological classification within the ``disk galaxy" class.
	   \item S : Galaxies that have relatively small angular size\footnote{
	   \RP{The majority (about 87\%) of S type galaxies have \Suk{a} Kron radius smaller than 50 arcsec, while only a small fraction (about 11\%) of normal spiral galaxies (Sa/Sb/Sc/Sd) belongs to these small objects.}
	   }
	    and did not have unique features within the spiral galaxy classifications scheme.
	\end{enumerate}

		\subsection{Secondary Morphology}
			
        \begin{figure}[htp]
        \epsscale{1}
        \plotone{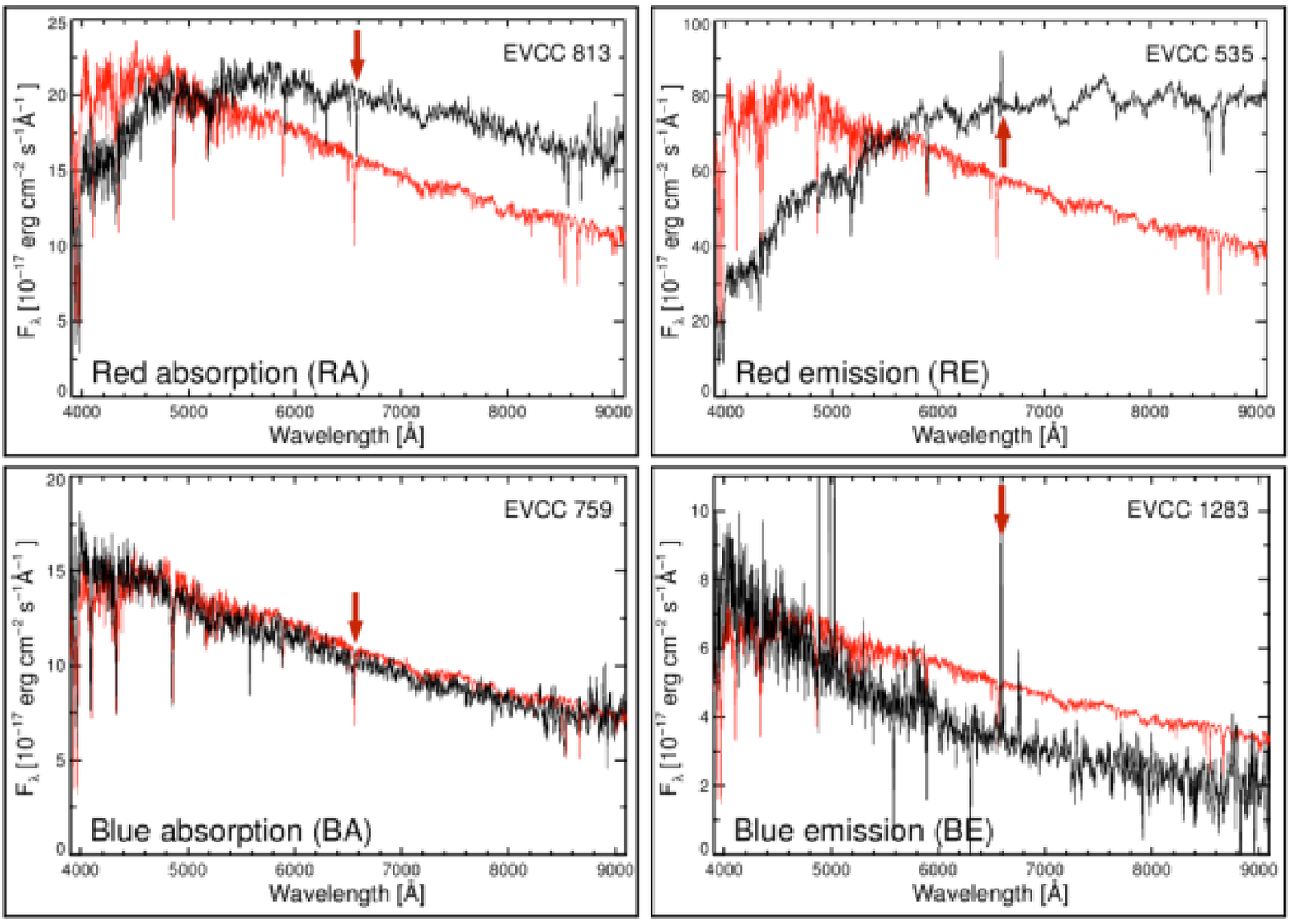}
		  \caption{
		  Examples of galaxies with various secondary morphologies. Red arrows indicate 
		  the H$\alpha$ line in emission or absorption.
		  In each panel, a model SED with age of 1 Gyr and solar metallicity from the population model is 
		  overplotted as a red solid curve. Secondary morphology is classified from comparison of this model SED with 
		  the observed spectrum of each galaxy (see main text for the details).
		  }
		  \label{Secon_mor_EVCC}
        \end{figure}

    The shape of the SED is related to the color of a galaxy while the stellar absorption and emission lines trace star formation activity within a galaxy \citep{Ken92}. In this context, we subdivide all galaxies into red and blue type galaxies based on their overall shape of the SEDs. We construct a model SED with age of 1 Gyr and solar metallicity from the population model (\citealt{Bru03}; see red solid curves in Figure ~\ref{Secon_mor_EVCC}) and compare it with \Suk{the} observed SED for each galaxy. When the wavelength of the peak intensity in the SED is shorter or longer compared to that of the model, we classify the galaxies as blue or red, respectively. We further subdivide the blue and red galaxies \Suk{into two subclasses based on the presence of Balmer absorption} and emission lines. Consequently, all galaxies included in the EVCC are classified as four types of secondary morphologies as follow (see Fig. ~\ref{Secon_mor_EVCC} for examples):  

	\begin{enumerate} [-]
    \item Red absorption (RA) galaxy : overall SED shape of a typical early-type red galaxy with H$\alpha$ absorption line.
    \item Red emission (RE) galaxy : overall SED shape of a typical early-type red galaxy, but has H$\alpha$ in emission.
    \item Blue absorption (BA) galaxy : overall SED shape of typical blue galaxy, and has H$\alpha$ absorption line.
    \item Blue emission (BE) galaxy : overall SED shape of typical blue galaxy with H$\alpha$ emission line. 
     In many cases, many emission lines (e.g., H$\alpha$, H$\beta$, and [OIII]$\lambda$5007) are prominent compared to the continuum.
    \end{enumerate}

    For galaxies with multiple SDSS spectra taken at various locations, we used the fiber spectrum obtained nearest to the photometric center of the galaxy (see also Sec. 2.3). We note that our secondary morphology generally provides spectral information of the central region of a galaxy due to the 3\,arcsec (i.e., 240 pc) fiber aperture of the SDSS spectrograph. Our secondary morphology classification is somewhat coarse, 
because galaxies belonging to the same secondary morphology can have a different SED slope and H$\alpha$ strength.
\\
	
    \subsection{Comparison Between Primary and Secondary Morphology}

        \begin{figure}
        \epsscale{1}
        \plotone{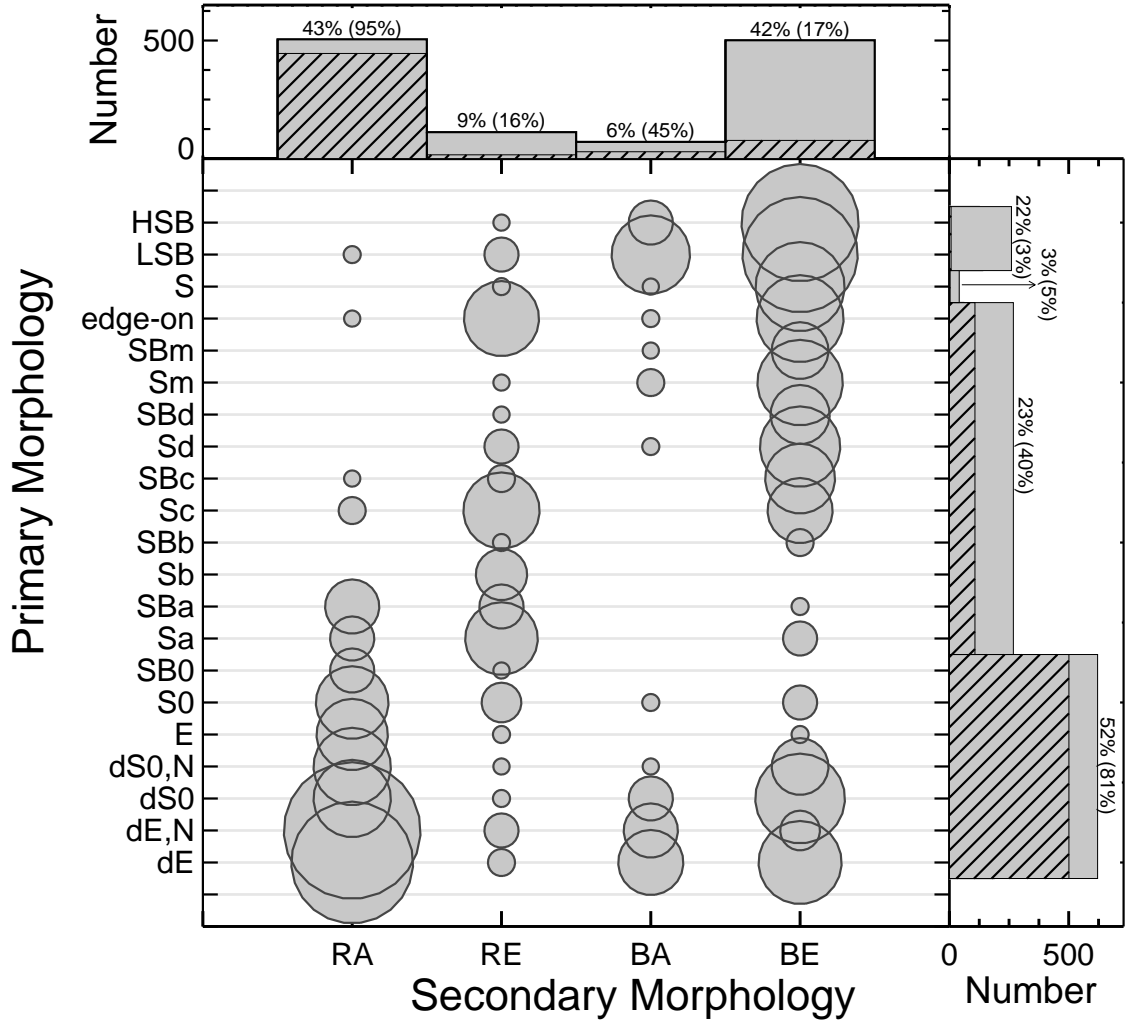}
        \caption{Comparison between primary and secondary morphology of 1189 galaxies in the EVCC. The symbol size is proportional to the number of galaxies. \Suk{In the upper and right panels}, grey filled histogram indicates the fraction of galaxies included in the secondary (RA, RE, BA, and BE) and primary morphology (early-type, spiral, and Irr), respectively. The hatched histograms \Suk{represent} the fraction of early-type galaxies (i.e., E/dE and S0/dS0) for each secondary morphology (upper panel) and that of red galaxies (i.e., RA and RE) for each primary morphology (right panel). The numbers above the histogram correspond to the fraction of morphology. 
\RP{The numbers in parentheses are the fraction of early-type and red (RA/RE) galaxies in each morphology bin, for the secondary and primary classification, respectively.}
        }		
        \label{Compar_mor_EVCC}
        \end{figure} 

    It has been known that the Hubble type of a galaxy closely correlates with its spectrum (\citealt{San10} and references therein). In Figure~\ref{Compar_mor_EVCC} we present a two-dimensional histogram comparing the two morphological classifications (primary vs. secondary) for the 1189 EVCC galaxies. We excluded 135 galaxies in the EVCC whose spectra are off-centered \Suk{by more than 5\,arcsec \SC{from the photometric galaxy center}.} Fig.~\ref{Compar_mor_EVCC} also presents the morphology distribution (filled histogram) as well as the percentage of early-type (i.e., E/dE and S0/dS0; hatched histogram of upper panel) and red (RA and RE; hatched histogram of right panel) galaxies in each morphology bin. We summarize the main features of comparison between primary and secondary morphology.
 
\begin{figure}[htp]
		\epsscale{0.8}
		\plotone{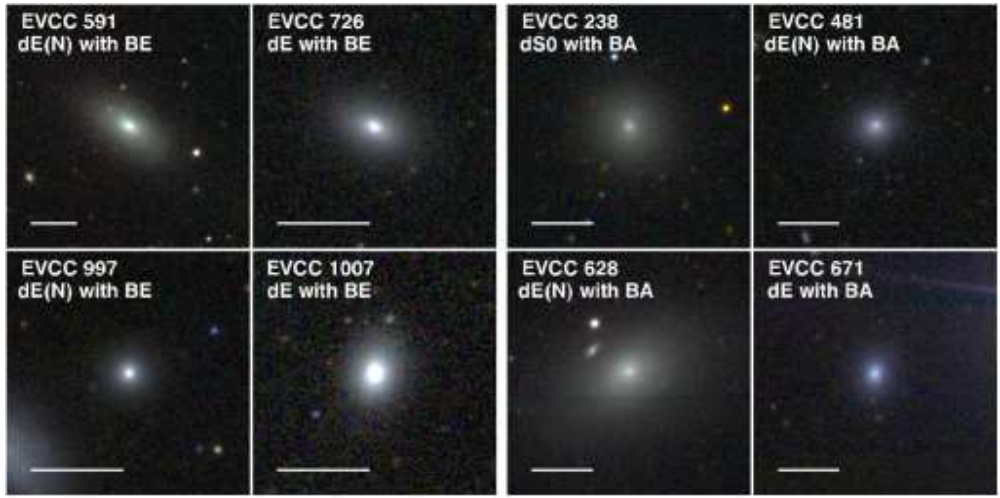}
		\caption{ \Suk{Examples of BE (left panels) and BA type (right panels) early-type dwarf galaxies that} exhibit blue core at their centers. The solid bar corresponds to 30 arcsec on the sky. North is up and east is to the left.
		}
		\label{dEBABE}
		\end{figure}

\begin{figure}[htp]
		\epsscale{0.8}
		\plotone{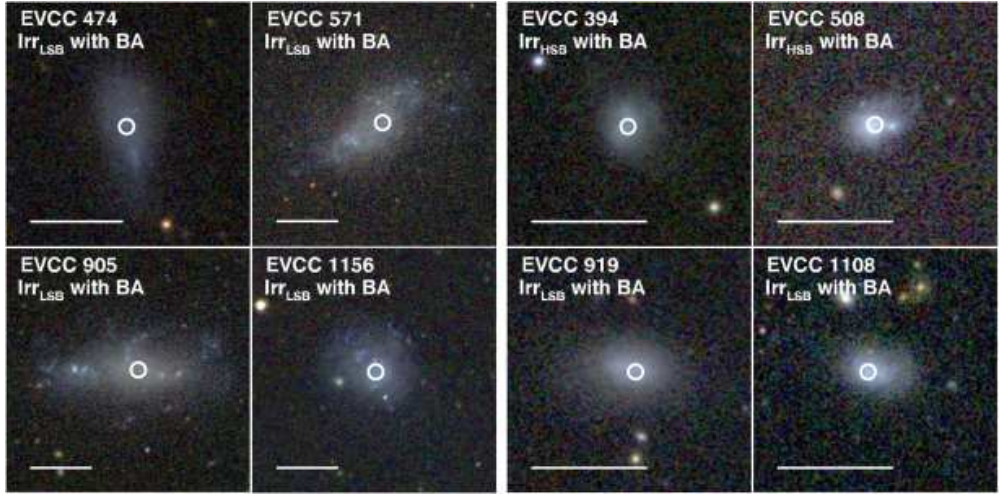}
		\caption{
		\SC{Examples of BA type irregular galaxies.} \Suk{The open circle indicates the fiber location} of the SDSS spectroscopic observation. The solid bar corresponds to 30 arcsec on the sky. North is up and east is to the left.
		}
		\label{IrrBA}
		\end{figure}

\begin{figure}[htp]
		\epsscale{0.8}
		\plotone{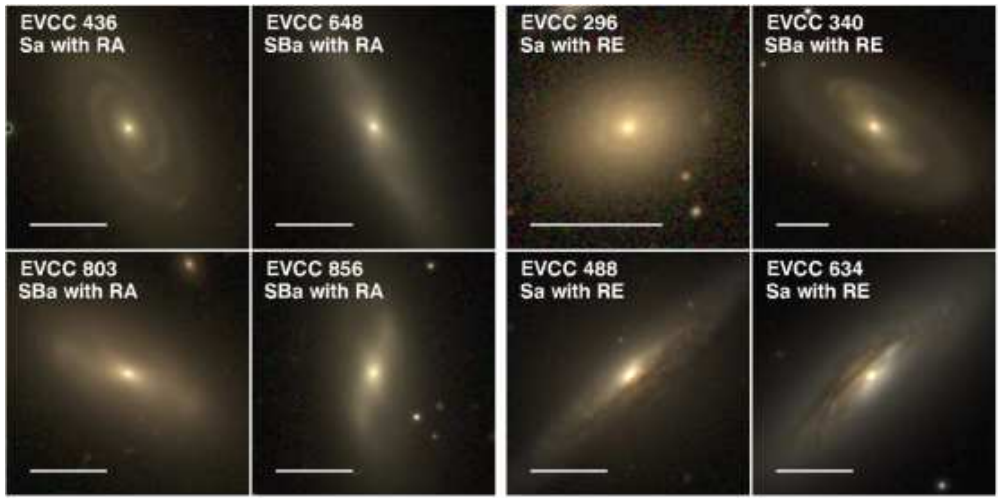}
		\caption{Examples of RA (left panels) and RE type (right panels) Sa and SBa galaxies. The solid bar corresponds to 30 arcsec on the sky. North is up and east is to the left.
		}
		\label{SaRARE}
		\end{figure}

\begin{figure}[htp]
		\epsscale{0.8}
		\plotone{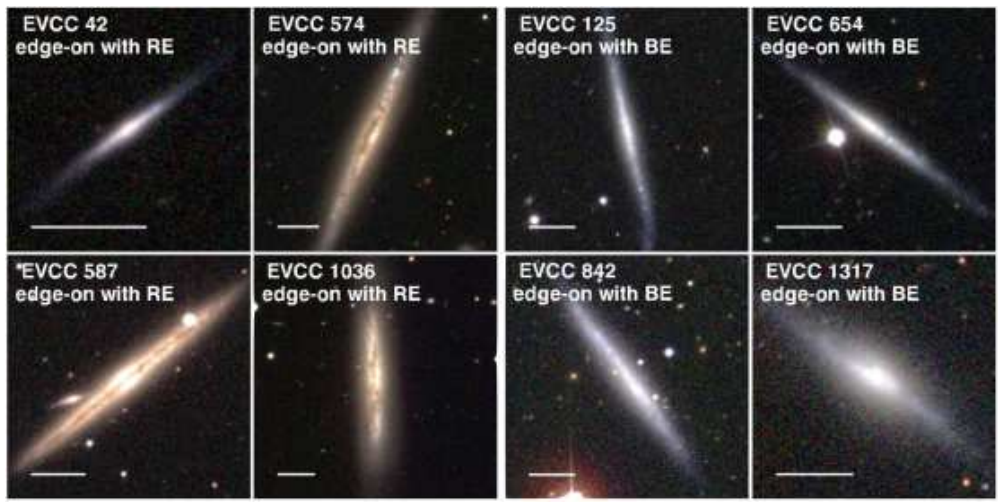}
		\caption{Examples of RE (left panels) and BE type (right panels) edge-on spiral galaxies. The solid bar corresponds to 30 arcsec on the sky. North is up and east is to the left.
		}
		\label{EdgeOnREBE}
		\end{figure}

    \begin{enumerate}[(1)]
    \item We find an expected general trend between photometric (i.e., primary) and spectroscopic (i.e., secondary) morphologies. \RP{Most (78\%)} early-type galaxies exhibit red spectra (i.e., RA type) and the majority \RP{(73\%)} of late-type galaxies show blue spectra (i.e., BE type). While our secondary morphology is obtained from the spectrum of the central region of the galaxy, it signifies that the secondary morphology also reflects global characteristics of the galaxies. However, the correlation between these two morphologies exhibits some scatter for both morphologies in the sense that there are \Suk{numerous} outliers in the main trend along the diagonal.
    \item The majority (95$\%$) of galaxies with red absorption line spectra are early-type galaxies (E/dE and S0/dS0). Only a small fraction (3$\%$) of RA type galaxies is bulge dominated Sa/SBa galaxies. There is a lack \Suk{\RP{(2\%)}} of late-type galaxies (Irrs and spiral galaxies later than Sa) within the RA type \citep[see also][]{San11}.
    \item While RE type galaxies extend over the entire range of primary morphology, they are dominated by spiral galaxies (mostly earlier than Sd galaxies) including edge-on galaxies. 
    \item Most BA type galaxies are early-type dwarf galaxies (dE and dS0) or HSB/LSB Irr galaxies. A small fraction (5$\%$) of early-type dwarf galaxies 
is of BA type. On the other hand, this type contains only a few spiral galaxies.
    \item BE type galaxies are mostly associated with the irregular galaxies and late spiral galaxies (Sc and Sd). Interestingly, a considerable 
fraction (14$\%$) of early-type dwarf galaxies (dE and dS0) is of BE type. Furthermore, some early-type dwarf galaxies classified as BA and BE type exhibit a blue core (see also Figure~\ref{dEBABE}; \citealt{Lis06}). 
    \item While the majority \RP{(86\%)} of irregular galaxies are of BE type, which is the result of ongoing star formation, a fraction (12$\%$) of them is classified as BA type with no prominent Balmer emission line. In the case of some BA type Irr galaxies with relatively large size and diffuse structure, SDSS fiber location, which is offset to the bright star\SC{-}forming knots in a galaxy, is responsible for their secondary morphology (see left panels of Figure ~\ref{IrrBA}). On the other hand, some relatively compact Irrs also show BA type, although the SDSS fiber is located at the galaxy center (see right panels of Fig. ~\ref{IrrBA}). These galaxies have probably experienced a recent quenching of star formation.     
     \item \RP{\Suk{A relatively dense obscuring band of interstellar dust is} often detectable in optical images of spiral galaxies. Different classes of dust lanes are defined depending on the morphology and inclination of galaxies (see \citealt{But13} for the details). A significant fraction (72\%, e.g., EVCC\,340, EVCC\,488, and EVCC\,634 in Figure~\ref{SaRARE}) of Sa and SBa galaxies in RE type shows dust lanes, while no such galaxy is seen in RA type.}
     \item The edge-on galaxies are dominated by RE and BE type ones. The BE type edge-on galaxies show overall bluer colors compared to RE type counterparts. Furthermore, a large fraction \RP{(75\%)} of RE type edge-on galaxies exhibits dust lanes \RP{(see EVCC\,574, EVCC\,587, and EVCC\,1036 in Figure~\ref{EdgeOnREBE}).}     
    \end{enumerate}

\begin{figure}
        \epsscale{1}
        \plotone{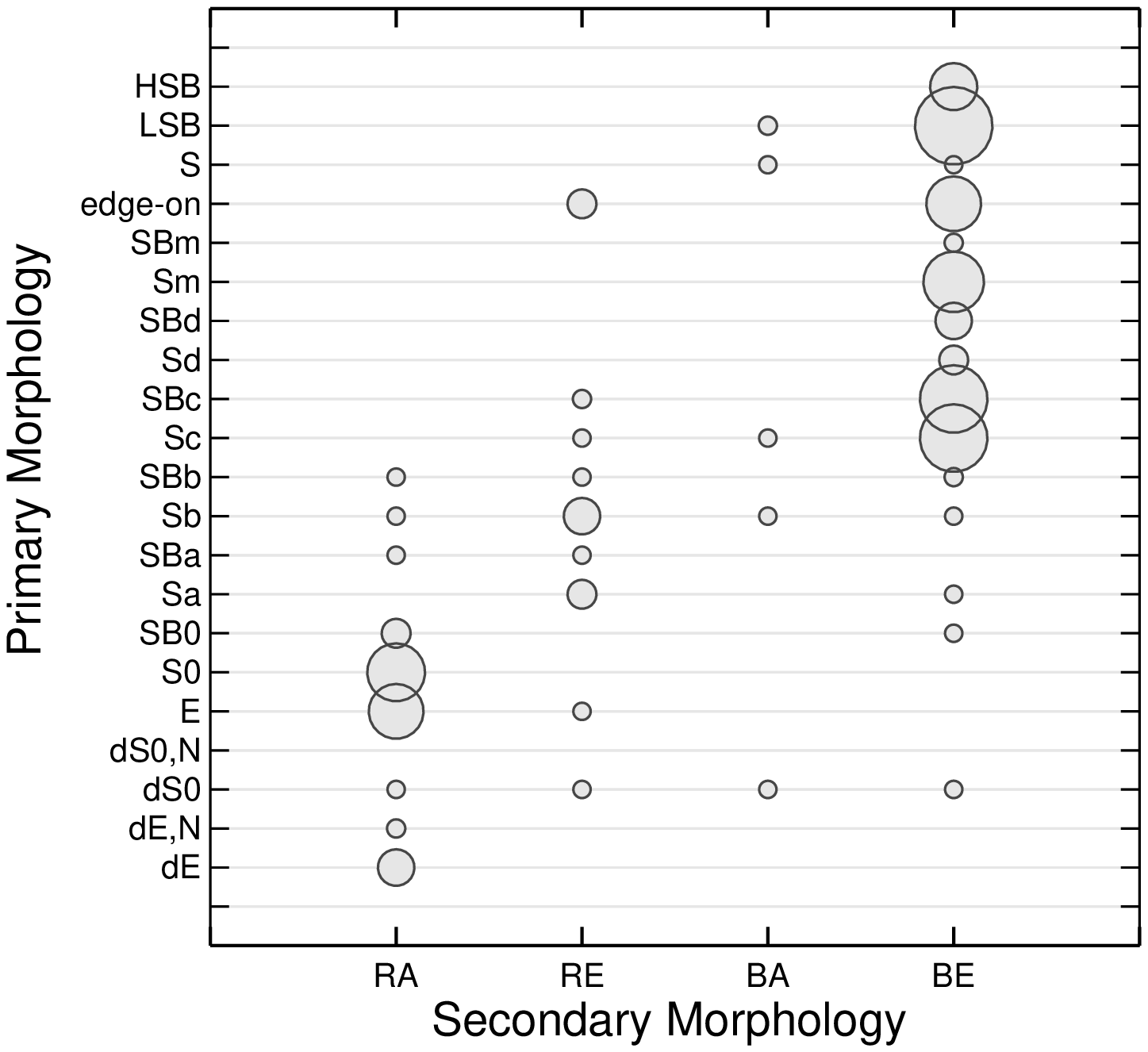}
        \caption{ Comparison between primary and secondary morphology of 135 galaxies with off-centered spectra in the EVCC. The fibers for these spectra are off-centered by more than 5\,arcsec from the photometric galaxy center. The symbol size is proportional to the number of galaxies.
        }		
        \label{Comp_mor_off}
        \end{figure}

    In Figure ~\ref{Comp_mor_off}, we also present a two dimensional histogram comparing \Suk{the} two morphological classifications for 135 galaxies whose spectra are off-centered by more than 5\,arcsec from the photometric galaxy center. This allows to examine whether \SC{off-center} spectra are likely to affect the secondary morphological classification. The first obvious \Suk{conclusion} drawn from inspection is that the overall distribution for the \SC{off-center} spectra appear to be quite consistent with that for \SC{on-center} spectra (see Fig.~\ref{Compar_mor_EVCC}). This denotes that, within our classification scheme, fiber location in the SDSS does not significantly affect the secondary morphology at a given primary morphology. However, note that, \Suk{in the case of spiral galaxies including edge-on systems} \SC{with off-center} spectra, they are mostly \RP{(72\%)} dominated by BE type and only a small fraction \RP{(20\%)} of RE type is found contrary to the cases of \SC{on-center} spectra. This is mostly due to that the fibers are positioned on the star\SC{-}forming regions in the disk rather than the bulges.

\section{PHOTOMETRY}
While the SDSS photometric pipeline is optimized for small, faint objects, the photometric parameters for large, bright galaxies are often affected by deblending (or shredding) and imperfect sky subtraction \citep{Aba04,Aba09,Wes10,Bla11}. Galaxies with large angular sizes, irregular morphology, and bright sub-clumps (e.g., HII regions and spiral arms) are often shredded into multiple, separate objects and thus the derived total flux for these galaxies are often unreliable. Furthermore, the SDSS photometric pipeline systematically underestimates the luminosities of galaxies with large size and/or in crowded fields, due to the overestimation of the sky background as they occupy a considerable fraction of the 256 $\times$ 256 pixel mask. \Suk{In order to overcome these problems}, we performed our own photometry for all galaxies in the EVCC.

The photometry of all EVCC galaxies \Suk{was performed using Source Extractor} \citep[SExtractor,][]{Ber96}. Additionally, \Suk{photometric parameters for} VCC galaxies that are not in common with the EVCC \Suk{were} also \Suk{measured}. For each galaxy, we created SDSS postage images in five passbands ($u, g, r, i,$ and $z$) depending on the angular size of the galaxy. For the construction of the background map, unlike the pipeline of the SDSS DR7, we masked out all proper objects in the image \citep[see also][]{Bla11}. This allows to eliminate any contribution of light from stars and background galaxies and then avoid over-estimation of sky background regardless of the BACK$\_$SIZE in SExtractor. In this masking image, masked regions which are defined by SExtractor segmentation image are filled by median value calculated from surrounding all pixels. Then, SExtractor constructs a background map by dividing the whole masking image into a grid of background meshes set by BACK$\_$SIZE = 256 pixels.

        \begin{figure}
        \epsscale{1}
        \begin{center}
        \plotone{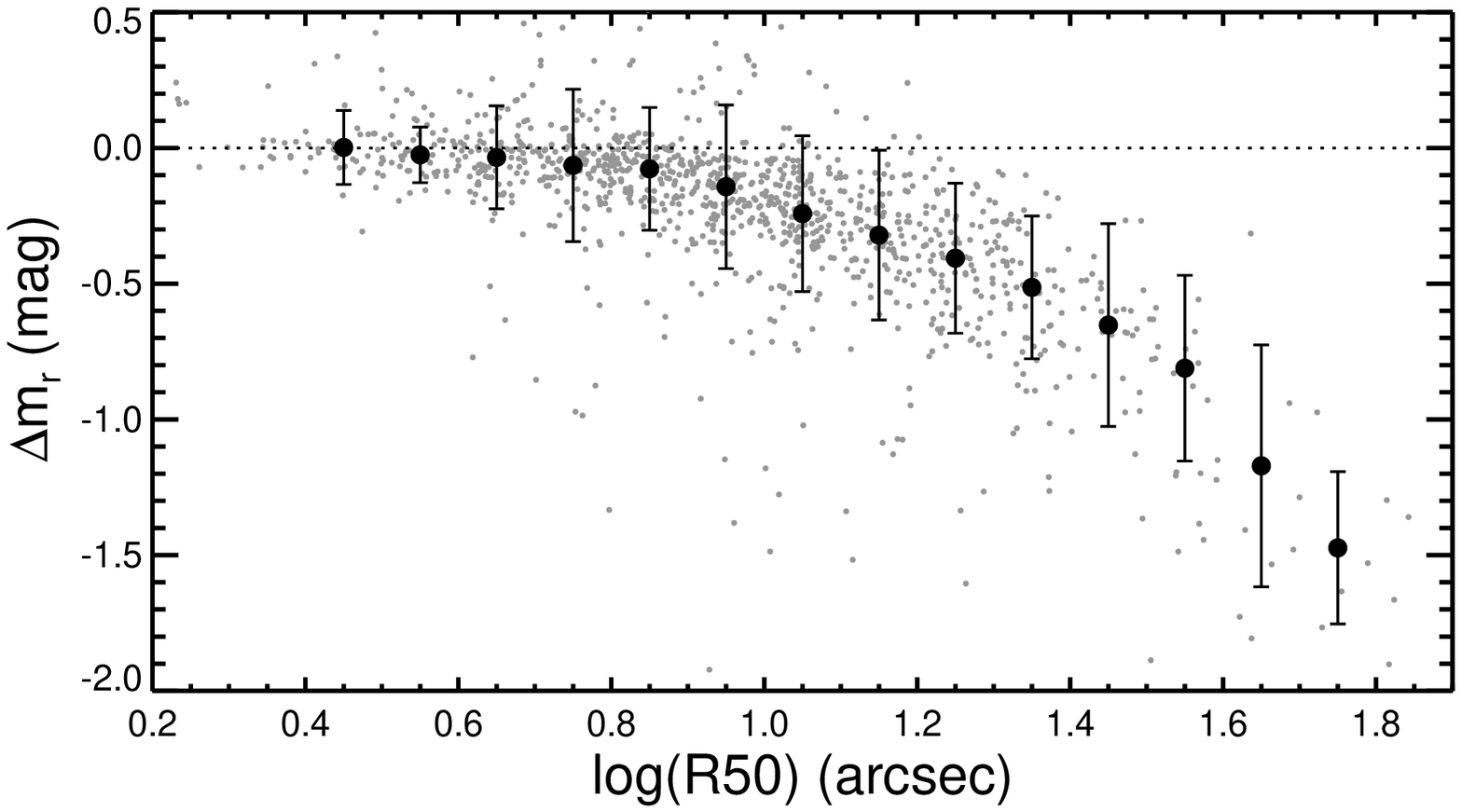}
        \caption{
        \RP{Difference between the galaxy's total $r$-band magnitude from our own photometry and Petrosian magnitude of the SDSS pipeline as a function of $r$-band half-light radius (R50).}
Filled circles and error bars denote the median value and standard deviation for every 0.1\,dex interval. 
        }
        \label{difSDSS}
        \end{center}
        \end{figure}

        \begin{figure}
        \epsscale{0.8}
        \begin{center}
        \plotone{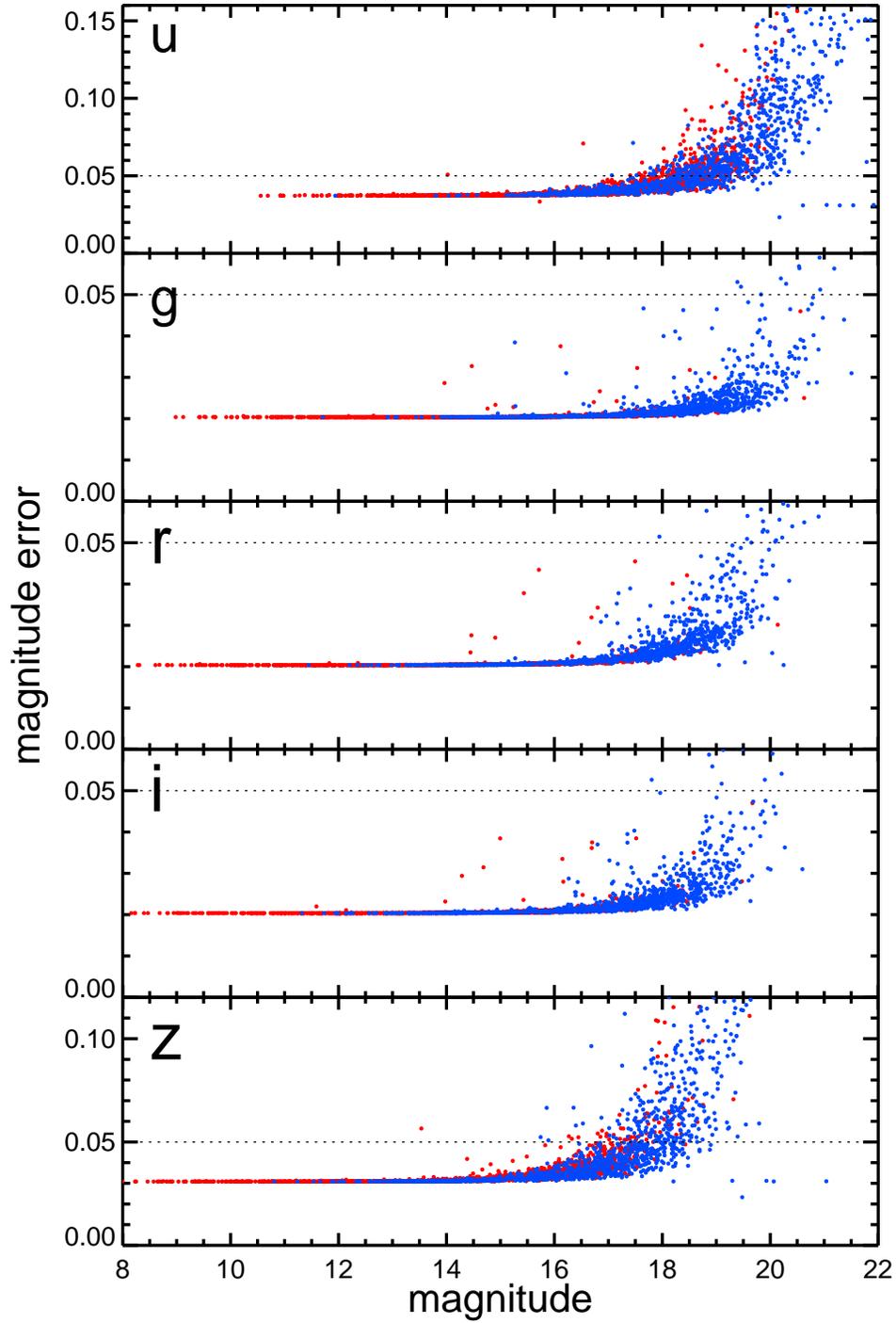}
        \caption{Magnitude error versus total magnitude in each passband. 
        Red and blue dots are for galaxies in the EVCC and included only 
        in the VCC, respectively. In each panel, the dotted line indicates the magnitude error of 0.05.}
        \label{magerr}
        \end{center}
        \end{figure}

Source detection and deblending were carried out on the $r$-band image, because it offers the highest S/N for most sources. The resulting positional and basic shape parameters were used for subsequent photometry in the other passbands. We found that the following parameters led to optimal source detection; DETECT$\_$THRESHOLD = 1 and DETECT$\_$MINAREA = 10. In some cases where galaxies were faint or affected by neighboring bright sources, different parameters were chosen (DETECT$\_$THRESHOLD = 0.5\,\Suk{$-$}\,1 and DETECT$\_$MINAREA = 5\,\Suk{$-$}\,50), which were verified by visual inspection, confirming that obvious galaxies are detected and large number of spurious detections are avoided. We selected individual values for the deblending parameters (DEBLEND$\_$MINCONT and DEBLEND$\_$NTHRESH) for each galaxy that minimize the fraction of shredded, spurious objects from target galaxy. The final values of parameters are selected by visual inspection of detected objects iteratively and typical values are DEBLEND$\_$MINCONT = 0.001\,\Suk{$-$}\,0.05 and DEBLEND$\_$NTHRESH = 1\,\Suk{$-$}\,64. If the deblending algorithm did not extract possible neighbor sources, we selected those by visual inspection and removed them in the image manually using the $IMEDIT$ command in $IRAF$.

We adopted MAG$\_$AUTO as the total magnitude of a galaxy, which provides the flux within $k$ (Kron factor) times the Kron radius. 
For the total flux, we used $k=2.5$, which is expected to recover more than 94$\%$ of the total galaxy light \citep{Ber96}. Once a flux was calculated, we converted all fluxes to asinh magnitudes \citep{Lup99}. 
In Figure~\ref{difSDSS}, \RP{we compare our own photometry of all galaxies in the EVCC with the petrosian magnitude from the SDSS photometric pipeline.} 
The main trend is that the SDSS pipeline systematically underestimates the fluxes of larger galaxies. \Suk{There is a difference of up to 1.5 magnitude.} This confirms that the photometry from the SDSS pipeline has problems with deblending and sky subtraction for large, bright galaxies.

Foreground Galactic extinction correction for each galaxy was then applied \citep{Sch98}. We used the reddening law of \citet{Car89} to derive the following: R$_{u}$ = 5.049, R$_{g}$ = 3.889, R$_{r}$ = 2.822, R$_{i}$ = 2.176, and R$_{z}$ = 1.567. 
\RP {Magnitude errors were estimated following the equation given by \citet{Lis08} where several different sources of uncertainties are added in quadrature: S/N uncertainty, sky level uncertainty, uncertainty in the determination of semimajor axis of aperture, photometric calibration uncertainty, and $u$-band leak uncertainty (see Sec. 4 of \citealt{Lis08} for the details). As an example, uncertainty of flux measurement for the $u$-band is given below:}

\begin{equation}
\Delta f_{\rm u}
= 
\Bigg(
\bigg(\frac{\sigma_{\rm u}\cdot \sqrt{N_{\rm
            pix}}}{f_{\rm u}} \bigg) ^2
+
2\cdot \bigg(\frac{0.002\,\sigma_{\rm u}\cdot N_{\rm pix}}{f_{\rm u}}\bigg)^2
+\nonumber\\
(10^{0.4\cdot 0.03}-1)^2
+
(10^{0.4\cdot 0.02}-1)^2
\Bigg)^{0.5}
\end{equation}

\RP{Here, $f_{\rm u}$ denotes the flux measurement. $\sigma_{\rm u}$ and $N_{\rm pix}$ \Suk{are the noise level per pixel and} the number of pixels included in the given aperture, respectively. The first term is the S/N uncertainty and the second term is the combination of sky level uncertainty and aperture uncertainty. The last two terms are the calibration uncertainty and the $u$-band leak uncertainty, respectively.} 
\Suk{Figure~\ref{magerr} presents the distribution of magnitude error for each passband.}

\section{THE CATALOG}
The EVCC contains 1589 galaxies. The 23 columns in Table 2 presents the full EVCC catalog including fundamental information such as membership, morphology, and photometric parameters:

\begin{enumerate}[Column 1:]
\setlength{\itemsep}{0pt}
\setlength{\parskip}{0pt}
\item EVCC ID number. Galaxies with numbers larger than 2000 have radial velocities from the NED. ID is marked with the right ascension.
\item VCC ID number.
\item NGC designation.
\item Right ascension (J2000) from SExtractor in degrees.
\item Declination (J2000) from SExtractor in degrees.
\item Right ascension (J2000) of the SDSS fiber location in degrees.
\item Declination (J2000) of the SDSS fiber location in degrees.
\item Offset between the photometric center and SDSS fiber location in arcsecs.
\item Heliocentric radial velocity from the SDSS in km\,s$^{-1}$.
\item Heliocentric radial velocity from the NED in km\,s$^{-1}$.
\item Membership of galaxy as derived from the Virgo cluster infall model (M = certain member, P = possible member, see Sec. 2.4).
\item Membership of galaxy from the VCC (M = member, P = possible member, B = background galaxy).
\item Primary morphology.
\item Secondary morphology.
\item Primary morphology in numerical type (see Table 1).
\item Morphology from the VCC.
\item u-band AUTO magnitude from SExtractor and error.
\item g-band AUTO magnitude from SExtractor and error.
\item r-band AUTO magnitude from SExtractor and error.
\item i-band AUTO magnitude from SExtractor and error.
\item z-band AUTO magnitude from SExtractor and error.
\item r-band Kron radius in arcsecs.
\item r-band half-light radius (R50) in arcsecs from SExtractor which encloses 50\% of the flux.\\
\end{enumerate}
In addition, we present in Table 3 the photometric parameters for \Suk{1183} galaxies in the VCC which are not in common with the EVCC.
They either have no available radial velocity information or have radial velocity with $cz$ $>$ 3,000\,km\,s$^{-1}$ from the SDSS and NED. 
The columns in Table 3 are:

\begin{enumerate}[Column 1:]
\setlength{\itemsep}{0pt}
\setlength{\parskip}{0pt}
\item VCC ID number. 
\item NGC designation.
\item Right ascension (J2000) from SExtractor in degrees.
\item Declination (J2000) from SExtractor in degrees.
\item Heliocentric radial velocity from the SDSS in km\,s$^{-1}$.
\item Heliocentric radial velocity from the NED in km\,s$^{-1}$.
\item Membership of galaxy from the VCC (M = member, P = possible member, B = background galaxy).
\item Morphology from the VCC.
\item u-band AUTO magnitude from SExtractor and error.
\item g-band AUTO magnitude from SExtractor and error.
\item r-band AUTO magnitude from SExtractor and error.
\item i-band AUTO magnitude from SExtractor and error.
\item z-band AUTO magnitude from SExtractor and error.
\item r-band Kron radius in arcsecs.
\item r-band half-light radius (R50) in arcsecs from SExtractor which encloses 50\% of the flux.
\end{enumerate}

\clearpage
\begin{flushleft}
\begin{deluxetable}{cccccccccccccccccccccccc}
\rotate
\setlength{\tabcolsep}{0.03in}
\tabletypesize{\tiny}
\tablecolumns{24}
\tablewidth{0pc}
\tablecaption{The Extended Virgo Cluster Catalog\tablenotemark{*}}
\tablehead{
\colhead{ID} & \colhead{ID} & \colhead{NGC} & \colhead{R.A.} & \colhead{Decl.} & \colhead{R.A.} & \colhead{Decl.} &\colhead{$\Delta$} & \colhead{cz} & \colhead{cz} & \colhead{Memb} &\colhead{Memb} & \colhead{Morp} &\colhead{Morp} &\colhead{Morp} & \colhead{Morp} & \colhead{u} & \colhead{g} & \colhead{r} & \colhead{i} & \colhead{z} & \colhead{R$_{Kron}$} & \colhead{R${50}$} \\

\colhead{EVCC} & \colhead{VCC} & \colhead{} & \colhead{} & \colhead{} & \colhead{Fiber} & \colhead{Fiber} &\colhead{} & \colhead{SDSS} & \colhead{NED} & \colhead{EVCC} & \colhead{VCC} & \colhead{Primary} & \colhead{Secondary}  & \colhead{index} & \colhead{VCC} & \colhead{AUTO} & \colhead{AUTO} & \colhead{AUTO} & \colhead{AUTO} & \colhead{AUTO} & \colhead{} \\

\colhead{ } & \colhead{ } & \colhead{} & \colhead{(deg)} & \colhead{(deg)} & \colhead{(deg)} & \colhead{(deg)} &\colhead{(arcsec)} & \colhead{(km\,s$^{-1}$)} & \colhead{(km\,s$^{-1}$)} & \colhead{} & \colhead{} & \colhead{} & \colhead{}  & \colhead{} & \colhead{} & \colhead{(mag)} & \colhead{(mag)} & \colhead{(mag)} & \colhead{(mag)} & \colhead{(mag)} & \colhead{(arcsec)}  & \colhead{(arcsec)}\\
\colhead{(1)} & \colhead{(2)} & \colhead{(3)} & \colhead{(4)} & \colhead{(5)} & \colhead{(6)} & \colhead{(7)} & \colhead{(8)} & \colhead{(9)} & \colhead{(10)} & \colhead{(11)} & \colhead{(12)} & \colhead{(13)} & \colhead{(14)} & \colhead{(15)} & \colhead{(16)} & \colhead{(17)} & \colhead{(18)} & \colhead{(19)} & \colhead{(20)} & \colhead{(21)} & \colhead{(22)} & \colhead{(23)}}     
\startdata
0001 & - & - & 175.0771 &  9.0099 & 175.0771 &  9.0099 &  0.241 & 1843.7 & 1860 & P & - & Sa       & RE & 201 & - & 14.95$\pm$0.037 & 13.50$\pm$0.020 & 12.75$\pm$0.020 & 12.33$\pm$0.020 & 12.01$\pm$0.030 &  50.01 & 22.44 &\\
0002 & - & - & 175.2361 & 14.0744 & 175.2364 & 14.0742 &  1.346 &  977.9 & - & M & - & S        & BE & 206 & - & 17.32$\pm$0.039 & 16.15$\pm$0.020 & 15.81$\pm$0.020 & 15.49$\pm$0.020 & 15.13$\pm$0.032 &  36.20 & 19.15 &\\
0003 & - & 3801 & 175.2447 & 11.4710 & 175.2448 & 11.4711 &  0.477 &  999.3 & 992 & M & - & Sc       & RE & 203 & - & 12.13$\pm$0.038 & 11.04$\pm$0.020 & 10.50$\pm$0.020 & 10.24$\pm$0.020 & 10.01$\pm$0.031 & 135.68 & 85.93 &\\
0004 & - & - & 175.4609 & 15.9736 & 175.4610 & 15.9737 &  0.468 &  753.1 & 748 & M & - & Irr(HSB) & BE & 310 & - & 15.50$\pm$0.037 & 14.59$\pm$0.020 & 14.12$\pm$0.020 & 13.90$\pm$0.020 & 13.76$\pm$0.030 &  26.01 & 14.96 &\\
0005 & - & - & 175.5747 & 14.9948 & 175.5792 & 14.9962 & 16.601 & 1021.5 & 1024 & M & - & Irr(LSB) & BE & 300 & - & 16.33$\pm$0.038 & 15.54$\pm$0.020 & 15.41$\pm$0.020 & 15.08$\pm$0.020 & 15.52$\pm$0.036 & 106.17 & 56.61 &\\
0006 & - & - & 175.6223 & 18.3326 & 175.6227 & 18.3325 &  1.503 &  906.4 & 922 & M & - & Sc       & BE & 203 & - & 14.14$\pm$0.037 & 13.32$\pm$0.020 & 12.99$\pm$0.020 & 12.83$\pm$0.020 & 12.92$\pm$0.031 &  99.26 & 53.93 &\\
0007 & - & - & 175.8163 & -1.3950 & 175.8171 & -1.3946 &  3.218 & 1891.6 & - & P & - & Irr(LSB) & BE & 300 & - & 17.35$\pm$0.040 & 16.83$\pm$0.020 & 16.56$\pm$0.021 & 16.36$\pm$0.021 & 16.26$\pm$0.035 &  21.04 & 13.11 &\\
0008 & - & - & 175.8624 & 11.3985 & 175.8624 & 11.3984 &  0.315 &  901.0 & - & M & - & Irr(HSB) & BE & 310 & - & 17.05$\pm$0.038 & 16.26$\pm$0.020 & 15.96$\pm$0.020 & 15.81$\pm$0.020 & 15.79$\pm$0.032 &  17.41 & 12.17 &\\
0009 & - & - & 175.9422 & 13.7076 & 175.9421 & 13.7075 &  0.238 & 2924.3 & 2920 & P & - & S0       & BE & 200 & - & 17.76$\pm$0.040 & 16.46$\pm$0.020 & 15.74$\pm$0.020 & 15.43$\pm$0.020 & 15.23$\pm$0.031 &  13.92 &  7.13 &\\
0010 & - & - & 176.0439 & -3.6689 & 176.0441 & -3.6686 &  1.303 & 1029.1 & - & P & - & Irr(LSB) & BE & 300 & - & 18.37$\pm$0.045 & 17.28$\pm$0.020 & 16.99$\pm$0.021 & 16.98$\pm$0.022 & 17.09$\pm$0.053 &  28.26 & 14.60 &\\
0011 & - & - & 176.1337 &  6.7041 & 176.1336 &  6.7040 &  0.522 & 1455.2 & - & P & - & Irr(HSB) & BE & 310 & - & 18.41$\pm$0.044 & 17.49$\pm$0.021 & 17.20$\pm$0.021 & 16.97$\pm$0.021 & 16.99$\pm$0.040 &  21.87 & 11.51 &\\
0012 & - & - & 176.1515 & 19.4781 & 176.1516 & 19.4776 &  1.844 & 2775.5 & - & P & - & Irr(LSB) & RE & 300 & - & 18.43$\pm$0.042 & 17.63$\pm$0.020 & 17.20$\pm$0.020 & 17.20$\pm$0.021 & 17.05$\pm$0.044 &  18.23 & 20.21 &\\
0013 & - & - & 176.1697 & 16.8995 & 176.1696 & 16.8996 &  0.474 &  902.1 & - & M & - & dS0      & BE & 410 & - & 18.40$\pm$0.054 & 16.72$\pm$0.020 & 16.25$\pm$0.020 & 16.11$\pm$0.020 & 15.37$\pm$0.032 &  16.30 & 10.91 &\\
0014 & - & - & 176.1809 & 11.2071 & 176.1811 & 11.2073 &  0.836 & 2944.3 & - & P & - & Irr(LSB) & BE & 300 & - & 17.93$\pm$0.042 & 17.00$\pm$0.020 & 16.75$\pm$0.021 & 16.70$\pm$0.021 & 16.63$\pm$0.042 &  23.66 & 15.33 &\\
0015 & - & - & 176.3174 & 13.8726 & 176.3174 & 13.8725 &  0.454 & 2974.0 & - & P & - & dE       & RA & 400 & - & 19.37$\pm$0.055 & 18.20$\pm$0.022 & 17.50$\pm$0.022 & 17.23$\pm$0.021 & 16.94$\pm$0.034 &  13.37 &  9.14 &\\
0016 & - & 3876 & 176.3609 &  9.1610 & 176.3620 &  9.1637 & 10.258 & 2862.8 & 2892 & P & - & SBb      & BE & 212 & - & 14.48$\pm$0.037 & 13.52$\pm$0.020 & 13.19$\pm$0.020 & 12.98$\pm$0.074 & 12.83$\pm$0.030 &  42.55 & 25.34 &\\
0017 & - & - & 176.4739 &  9.8998 & 176.4740 &  9.8997 &  0.444 & 2918.5 & - & P & - & dS0      & BE & 410 & - & 18.51$\pm$0.044 & 17.26$\pm$0.020 & 16.93$\pm$0.021 & 16.76$\pm$0.021 & 16.50$\pm$0.038 &  13.72 &  7.88 &\\
0018 & - & - & 176.5169 & 11.5813 & 176.5168 & 11.5813 &  0.317 & 2970.2 & 2969 & P & - & S0       & BE & 200 & - & 16.47$\pm$0.037 & 15.47$\pm$0.020 & 15.00$\pm$0.020 & 14.75$\pm$0.020 & 14.61$\pm$0.031 &  17.35 &  8.74 &\\
0019 & - & - & 176.7524 & -0.2939 & 176.7530 & -0.2942 &  2.442 & 1463.1 & 1478 & P & - & dS0      & BE & 410 & - & 16.24$\pm$0.037 & 15.10$\pm$0.020 & 14.60$\pm$0.020 & 14.34$\pm$0.020 & 14.21$\pm$0.031 &  33.48 & 17.23 &\\
0020 & - & - & 176.7792 &  3.1064 & 176.7791 &  3.1064 &  0.232 & 1012.7 & - & M & - & dE       & BE & 400 & - & 18.63$\pm$0.044 & 17.69$\pm$0.020 & 17.36$\pm$0.021 & 17.06$\pm$0.021 & 16.72$\pm$0.035 &  17.45 & 11.48 &\\
0021 & - & - & 176.8230 & 16.3346 & 176.8232 & 16.3345 &  0.593 &  809.0 & - & M & - & dE       & BE & 400 & - & 18.53$\pm$0.045 & 17.70$\pm$0.021 & 17.29$\pm$0.022 & 16.96$\pm$0.021 & 17.09$\pm$0.039 &  16.99 & 10.04 &\\
0022 & - & - & 177.0440 & -1.9892 & 177.0443 & -1.9891 &  0.884 & 1529.3 & - & P & - & Irr(LSB) & BE & 300 & - & 17.54$\pm$0.040 & 16.81$\pm$0.020 & 16.64$\pm$0.021 & 16.65$\pm$0.021 & 16.46$\pm$0.039 &  24.45 & 16.41 &\\
0023 & - & - & 177.0683 & 18.6423 & 177.0680 & 18.6425 &  1.145 & 1010.6 & - & M & - & Irr(LSB) & BE & 300 & - & 17.13$\pm$0.039 & 16.37$\pm$0.020 & 16.11$\pm$0.020 & 15.89$\pm$0.020 & 15.52$\pm$0.033 &  30.68 & 20.85 &\\
0024 & - & - & 177.1789 & 17.1811 & 177.1797 & 17.1814 &  2.860 & 1058.7 & - & M & - & Irr(LSB) & BE & 300 & - & 17.89$\pm$0.043 & 17.11$\pm$0.020 & 16.80$\pm$0.021 & 16.60$\pm$0.021 & 16.83$\pm$0.041 &  27.44 & 17.21 &\\
0025 & - & - & 177.2101 & -2.0326 & 177.2100 & -2.0322 &  1.527 & 1712.5 & 1723 & P & - & Sc       & BE & 203 & - & 14.57$\pm$0.037 & 13.67$\pm$0.020 & 13.34$\pm$0.020 & 12.91$\pm$0.020 & 13.02$\pm$0.031 & 184.37 & 89.82 &\\
\enddata
\tablecomments{
Table 2 is presented in its entirety in the electronic edition of the Astrophysical Journal Supplement. A portion is shown here for guidance regarding its form and content.
}
\tablenotetext{*}{The catalog of the EVCC will be updated at our own website after the publication.}

\end{deluxetable}
\end{flushleft}

\begin{flushleft}
\begin{deluxetable}{ccccccccccccccccc}
\tablecaption{Galaxies in the VCC\tablenotemark{*}}
\rotate
\setlength{\tabcolsep}{0.06in}
\tabletypesize{\tiny}
\tablecolumns{17}
\tablewidth{0pc}
\tablehead{
\colhead{ID} & \colhead{NGC} & \colhead{R.A.} & \colhead{Decl.} &  \colhead{cz} & \colhead{cz} &\colhead{Memb} & \colhead{Morp} & \colhead{u} & \colhead{g} & \colhead{r} & \colhead{i} & \colhead{z} & \colhead{R$_{Kron}$} & \colhead{R${50}$} \\

\colhead{VCC} & \colhead{} & \colhead{} & \colhead{} & \colhead{SDSS} & \colhead{NED} & \colhead{VCC} & \colhead{VCC} & \colhead{AUTO}  & \colhead{AUTO} & \colhead{AUTO} & \colhead{AUTO} & \colhead{AUTO} & \colhead{} & \colhead{}\\

\colhead{} & \colhead{} & \colhead{(deg)} & \colhead{(deg)} & \colhead{(km\,s$^{-1}$)} & \colhead{(km\,s$^{-1}$)} & \colhead{} & \colhead{} & \colhead{(mag)} & \colhead{(mag)} & \colhead{(mag)} & \colhead{(mag)} & \colhead{(mag)} & \colhead{(arcsec)} & \colhead{(arcsec)} & \colhead{} \\

\colhead{(1)} & \colhead{(2)} & \colhead{(3)} & \colhead{(4)} & \colhead{(5)} & \colhead{(6)} & \colhead{(7)} & \colhead{(8)} & \colhead{(9)} & \colhead{(10)} & \colhead{(11)} & \colhead{(12)} & \colhead{(13)} & \colhead{(14)} & \colhead{(15)}}
\startdata
2 &     - &   182.1055 &    13.8284 &            - &            - &   M & dE2:              &   18.89$\pm$ 0.050 &   17.82$\pm$ 0.022 &   17.37$\pm$ 0.022 &   16.91$\pm$ 0.021 &   17.19$\pm$ 0.049 &    24.52 &     8.07&\\
   3 &     - &   182.1121 &    13.5238 &       6774.8 &       6805.8 &   P & BCD?              &   17.56$\pm$ 0.039 &   16.89$\pm$ 0.021 &   16.67$\pm$ 0.021 &   16.54$\pm$ 0.021 &   16.49$\pm$ 0.034 &    16.24 &     4.08&\\
   5 &     - &   182.1409 &    15.1188 &            - &            - &   M & dE4               &   19.01$\pm$ 0.054 &   18.09$\pm$ 0.021 &   17.31$\pm$ 0.022 &   16.76$\pm$ 0.022 &   17.42$\pm$ 0.071 &    26.12 &     7.38&\\
   6 &     - &   182.2152 &     9.1316 &       8172.0 &       8180.7 &   B & SBa               &   16.62$\pm$ 0.038 &   15.40$\pm$ 0.020 &   14.84$\pm$ 0.020 &   14.57$\pm$ 0.020 &   14.35$\pm$ 0.031 &    21.16 &     5.67&\\
   7 &     - &   182.3274 &    11.4301 &      18904.5 &      18900.3 &   B & SBc(s)I           &   17.04$\pm$ 0.040 &   15.75$\pm$ 0.021 &   15.18$\pm$ 0.021 &   14.84$\pm$ 0.021 &   14.61$\pm$ 0.031 &    22.73 &     7.64&\\
   8 &     - &   182.3373 &    13.5260 &            - &            - &   M & dE0               &   19.94$\pm$ 0.089 &   18.56$\pm$ 0.022 &   18.12$\pm$ 0.024 &   17.91$\pm$ 0.024 &   17.47$\pm$ 0.040 &    18.51 &     6.68&\\
  11 &     - &   182.3987 &     6.7425 &            - &            - &   M & dE6               &   18.21$\pm$ 0.050 &   16.75$\pm$ 0.021 &   16.48$\pm$ 0.021 &   16.02$\pm$ 0.021 &   16.29$\pm$ 0.037 &    50.96 &    13.88&\\
  12 &     - &   182.4348 &    12.1258 &       8643.3 &       8666.1 &   B & SBa(s)            &   16.96$\pm$ 0.039 &   15.41$\pm$ 0.020 &   14.64$\pm$ 0.020 &   14.25$\pm$ 0.020 &   13.96$\pm$ 0.031 &    25.50 &     4.39&\\
  13 &     - &   182.4411 &    13.5520 &            - &            - &   P & ?                 &   19.31$\pm$ 0.060 &   18.17$\pm$ 0.021 &   17.76$\pm$ 0.022 &   17.46$\pm$ 0.021 &   17.00$\pm$ 0.037 &    27.87 &     9.38&\\
  14 &     - &   182.4631 &    11.2567 &      17931.9 &      17930.1 &   P & BCD?              &   17.73$\pm$ 0.041 &   16.73$\pm$ 0.021 &   16.34$\pm$ 0.021 &   16.07$\pm$ 0.021 &   15.96$\pm$ 0.033 &    10.30 &     2.79&\\
  16 &     - &   182.5047 &    14.6155 &       6806.4 &       6774.6 &   P & ImIII?            &   17.92$\pm$ 0.043 &   16.97$\pm$ 0.021 &   16.58$\pm$ 0.021 &   16.42$\pm$ 0.022 &   16.40$\pm$ 0.043 &    23.12 &     6.81&\\
  18 &     - &   182.5496 &    12.3255 &       8755.7 &       8759.1 &   P & Sc(s)II           &   16.13$\pm$ 0.038 &   14.92$\pm$ 0.020 &   14.36$\pm$ 0.020 &   14.02$\pm$ 0.020 &   13.70$\pm$ 0.031 &    36.17 &     7.15&\\
  19 &     - &   182.5575 &    13.1880 &       6807.6 &       6807.6 &   P & BCD?              &   17.65$\pm$ 0.045 &   16.00$\pm$ 0.020 &   15.15$\pm$ 0.020 &   14.65$\pm$ 0.020 &   14.36$\pm$ 0.031 &    20.38 &     5.01&\\
  20 &     - &   182.5787 &    12.3297 &            - &            - &   P & ?                 &   18.91$\pm$ 0.059 &   17.85$\pm$ 0.021 &   17.42$\pm$ 0.022 &   16.97$\pm$ 0.021 &   16.72$\pm$ 0.037 &    20.09 &     5.78&\\
  23 &     - &   182.6040 &    13.3663 &            - &            - &   P & dE5?              &   20.22$\pm$ 0.138 &   19.04$\pm$ 0.026 &   17.88$\pm$ 0.023 &   17.97$\pm$ 0.024 &   18.29$\pm$ 0.065 &    22.80 &     7.23&\\
  27 &     - &   182.6744 &    13.3311 &       6828.6 &       6819.6 &   B & SBc(s)I           &   15.93$\pm$ 0.038 &   14.61$\pm$ 0.020 &   14.07$\pm$ 0.020 &   13.79$\pm$ 0.020 &   13.58$\pm$ 0.031 &    48.49 &     9.90&\\
  28 &     - &   182.6894 &    15.8651 &       6391.3 &       6329.4 &   B & SBc               &   16.78$\pm$ 0.040 &   15.71$\pm$ 0.020 &   15.29$\pm$ 0.021 &   15.17$\pm$ 0.021 &   15.11$\pm$ 0.032 &    33.47 &     8.96&\\
  35 &     - &   182.8312 &    11.9099 &            - &            - &   P & ?                 &   20.49$\pm$ 0.110 &   19.52$\pm$ 0.023 &   19.34$\pm$ 0.032 &   18.69$\pm$ 0.025 &   18.02$\pm$ 0.040 &     9.34 &     3.23&\\
  36 &     - &   182.8677 &    13.5838 &            - &            - &   M & dE0:              &   19.50$\pm$ 0.067 &   18.70$\pm$ 0.022 &   18.14$\pm$ 0.024 &   17.65$\pm$ 0.022 &   17.97$\pm$ 0.047 &    16.13 &     5.59&\\
  38 &     - &   182.9489 &    12.1436 &       4010.6 &       4022.7 &   B & Sc(s)II-III       &   15.42$\pm$ 0.038 &   14.24$\pm$ 0.020 &   13.70$\pm$ 0.020 &   13.39$\pm$ 0.020 &   13.18$\pm$ 0.031 &    63.68 &    15.58&\\
  39 &     - &   182.9974 &    15.4013 &       7111.2 &       7110.0 &   B & Sbc(on\_edge)      &   16.11$\pm$ 0.039 &   14.31$\pm$ 0.020 &   13.45$\pm$ 0.020 &   13.01$\pm$ 0.020 &   12.69$\pm$ 0.031 &    53.04 &    10.04&\\
  40 &     - &   183.0133 &    14.9046 &       6885.7 &       6953.7 &   B & Sbc(s)            &   16.83$\pm$ 0.040 &   15.73$\pm$ 0.020 &   15.20$\pm$ 0.021 &   14.93$\pm$ 0.021 &   14.73$\pm$ 0.031 &    29.45 &     9.48&\\
  42 &     - &   183.0252 &    14.9529 &            - &            - &   M & dE0               &   19.39$\pm$ 0.053 &   18.95$\pm$ 0.023 &   18.22$\pm$ 0.023 &   17.88$\pm$ 0.022 &   17.61$\pm$ 0.039 &    24.50 &     8.87&\\
  43 &  4164 &   183.0228 &    13.2056 &      19467.7 &      17512.2 &   P & E3                &   16.74$\pm$ 0.044 &   14.95$\pm$ 0.020 &   13.95$\pm$ 0.020 &   13.67$\pm$ 0.020 &   13.40$\pm$ 0.031 &    77.83 &    16.16&\\
\enddata
\tablecomments{
Table 3 is presented in its entirety in the electronic edition of the  Astrophysical Journal Supplement. A portion is shown here for guidance regarding its form and content.
}
\tablenotetext{*}{The VCC galaxies which are not in common with the EVCC. The catalog of the EVCC will be updated at our own website after the publication. }
\end{deluxetable}
\end{flushleft}

\section{Comparison between the EVCC and VCC}
In the following we compare the EVCC and VCC in terms of morphology, projected spatial distribution, and galaxy luminosity function. 
For that purpose we have arranged the morphologies of the galaxies in the two catalogs in a consistent way. VCC galaxies with mixed morphologies \citep[e.g., dE/dS0, dE/Im, Im/BCD, etc;][]{Bin85} have been assigned to the first morphology class. For VCC galaxies with uncertain morphologies, indicated by ``:" or ``?"  \citep[e.g., dE:, dE?, etc;][]{Bin85}, we take their morphologies at face value. The Im and BCD galaxies in the VCC approximately correspond to the Irr(LSB) and Irr(HSB) type galaxies in the EVCC, respectively. In addition, we do not discriminate between barred and non-barred galaxies in both cases of the VCC and EVCC (e.g., from dSB0 and SBc to dS0 and Sc, respectively).

    \subsection{Comparison of Morphology}
        \begin{figure*}
        \epsscale{1}
        \plotone{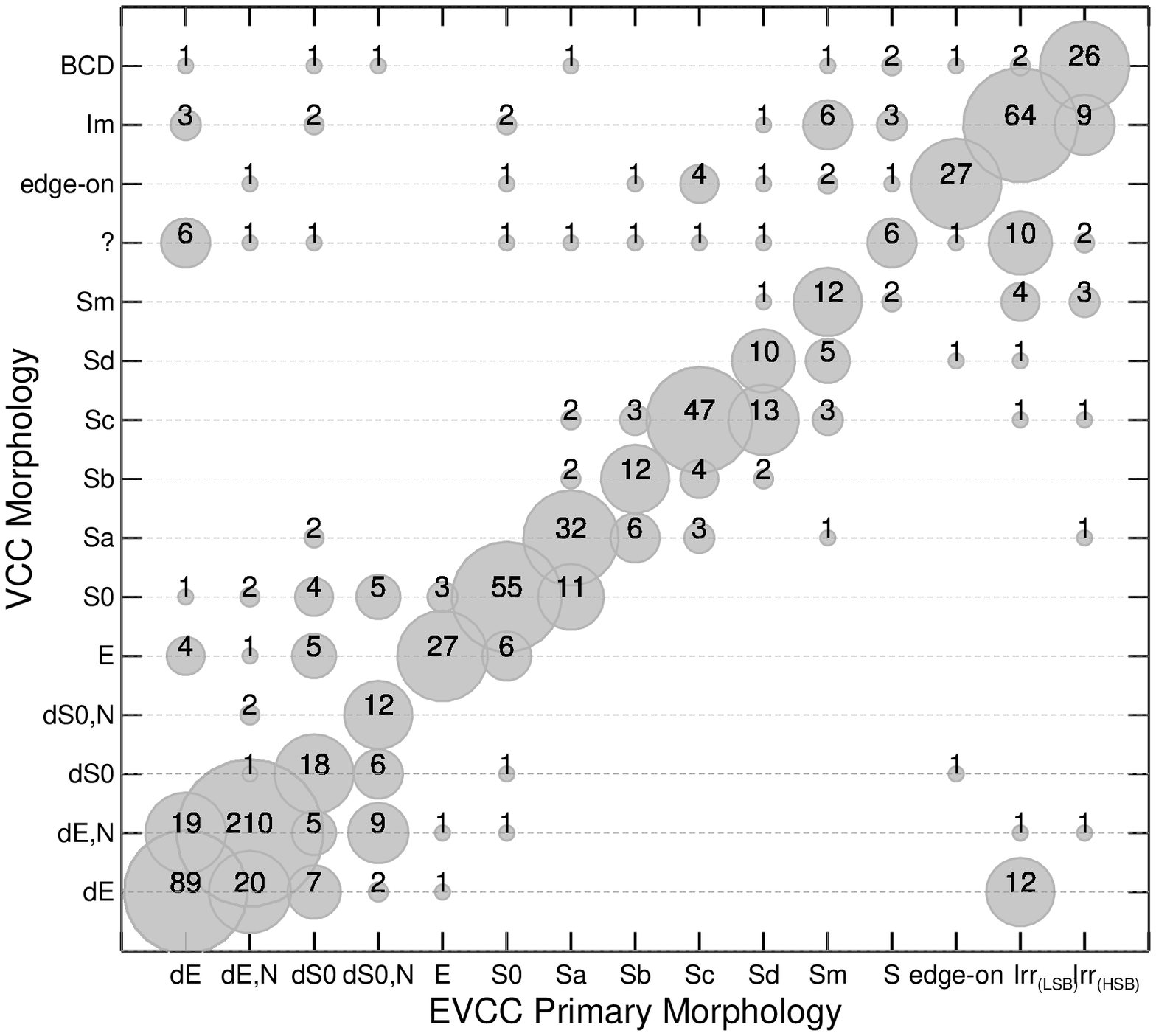}
        \caption{
        Comparison between EVCC primary morphological classification 
        and VCC classification for 913 objects in common. 
        The symbol size is proportional to the number of galaxies.
        }
        \label{ComparVCC}
        \end{figure*}
		
        \begin{figure*}
        \epsscale{0.8}
        \plotone{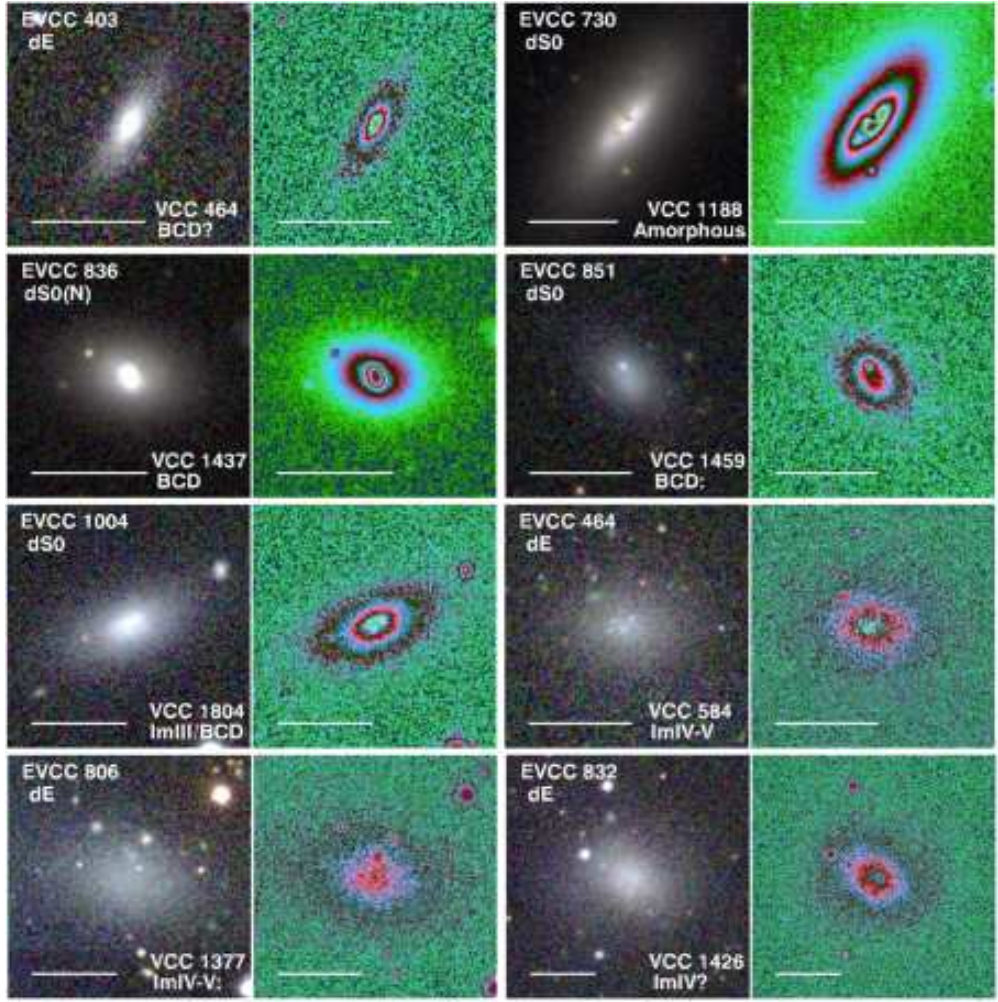}
        \caption{
        Examples of eight VCC galaxies with late-type morphology that were reclassified to early-type dwarf galaxies in the EVCC. The left and right panels show the SDSS $gri$ combined color and $r$-band images, respectively. The intensity is represented by alternating colors in the $r$-band image to highlight the shape of isophotes and substructures within a galaxy. The solid bar corresponds to 30\,arcsec on the sky. North is up and east is to the left.
        }
        \label{IrrtodE}
        \end{figure*}	
		
        \begin{figure*}[htp]
        \epsscale{1.05}
        \begin{center} 
        \plotone{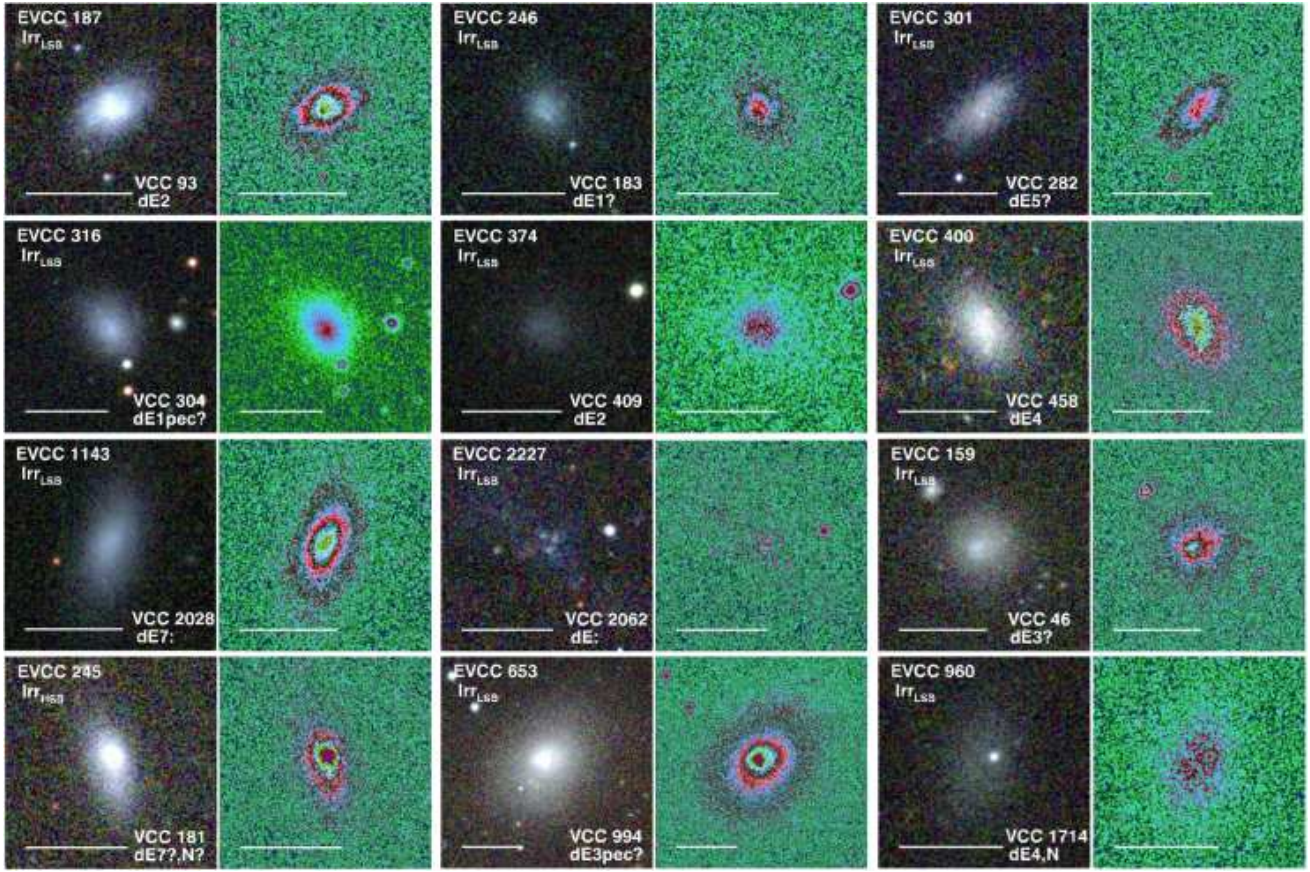}
        \caption{
        \Suk{\SC{Examples of twelve} VCC early-type dwarf galaxies that were reclassified to irregular galaxies in the EVCC. \SC{The left and right panels show the SDSS $gri$ combined color and $r$-band images, respectively. The intensity is represented by alternating colors in the $r$-band image to highlight the shape of isophotes and substructures within a galaxy. The solid bar corresponds to 30 arcsec on the sky. North is up and east is to the left.}}
        }		
        \label{dEtoIrr}
        \end{center}
        \end{figure*}		

        \begin{figure}[htp]
        \epsscale{0.8}
        \plotone{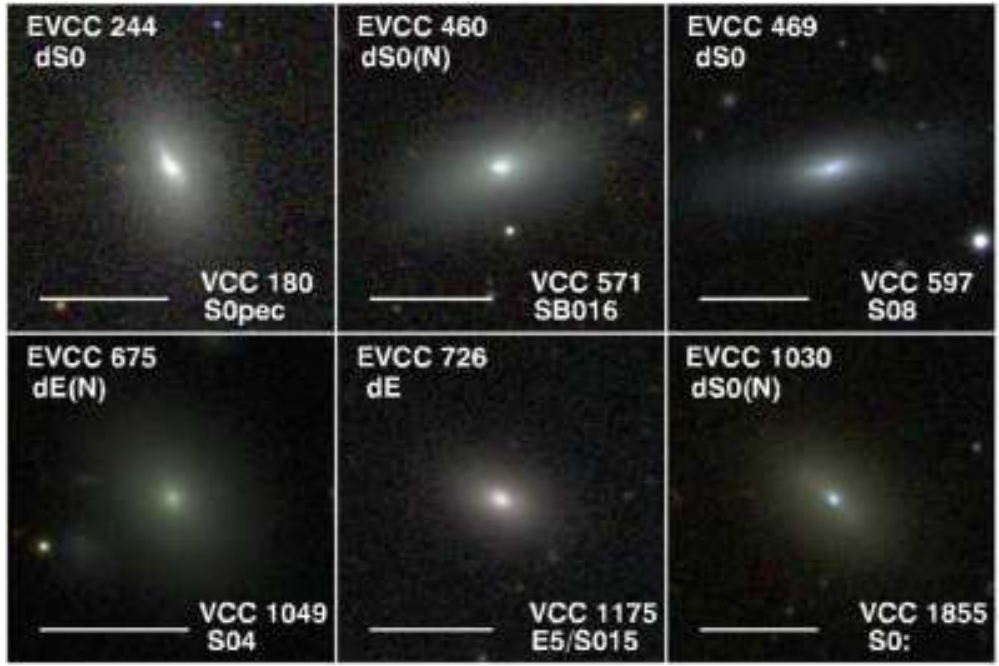}
        \caption{
        Examples of  reclassification to dE/dS0 galaxies in the EVCC from E/S0 morphologies in the VCC. The solid bar corresponds to 30\,arcsec on the sky. North is up and east is to the left.
        }
        \label{GToD}
        \end{figure}

        \begin{figure}[htp]
        \begin{center}
        \epsscale{0.8}
        \plotone{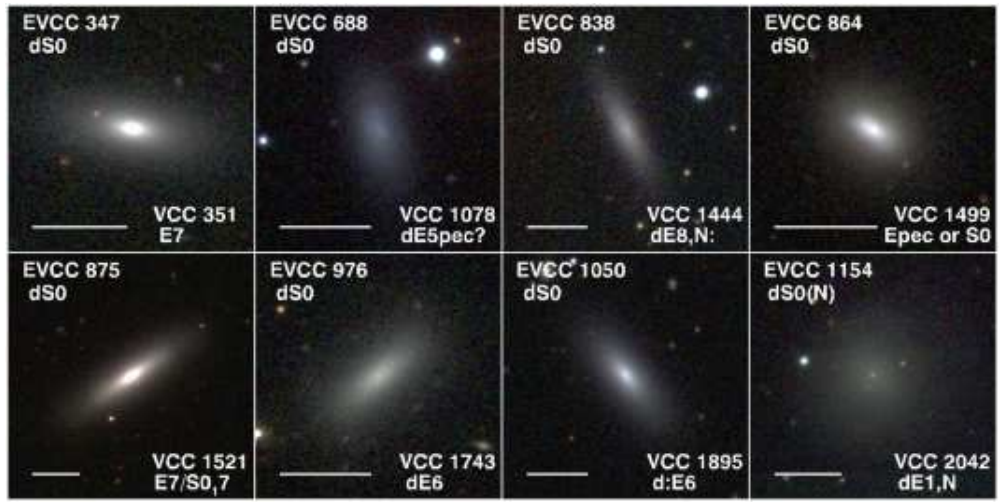}
        \caption{
        Examples of eight dwarf lenticular (dS0) galaxies in the EVCC with different morphology in the VCC. The solid bar corresponds to 30\,arcsec on the sky. North is up and east is to the left.}
        \label{LenVCC}
        \end{center}
        \end{figure}

        \begin{figure}[htp]
        \begin{center}
        \epsscale{0.8}
        \plotone{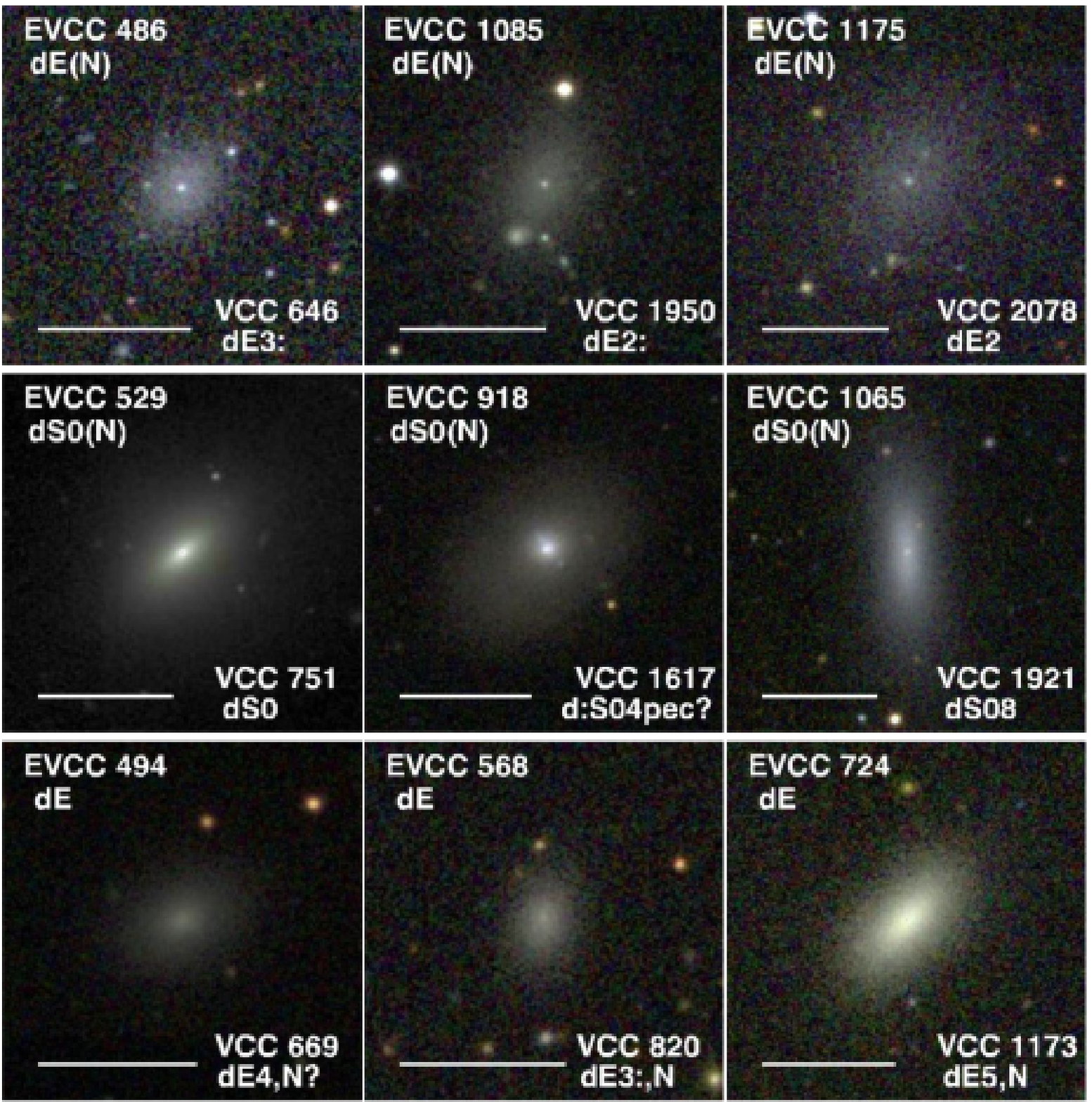}
        \caption{
        (Top and middle panels) Examples of SDSS images of dEs and dS0s where we found a nucleus. (Bottom panels) Examples of SDSS images of three dEs where a nucleus could not be confirmed. The solid bar corresponds to 30 arcsec on the sky. North is up and east is to the left.
        }
        \label{dEN}
        \end{center}
        \end{figure}

In Figure~\ref{ComparVCC}, we present a two-dimensional histogram comparing the morphologies for the 913 EVCC galaxies that are in common with the VCC. Our morphological classification in the EVCC is generally consistent with the VCC classification. There is a tight trend along the diagonal between two catalogs. Differences are typically at the level of one or two morphological subtypes. The morphological classifications of \Suk{90$\%$ (822 of 913)} galaxies are in good agreement within deviation of two morphological types.

However, it is worth to note that the relationship has some scatters at a fixed morphological type.  In the following we discuss the most interesting outliers, where the morphological classification between our EVCC and the VCC disagree more significantly.

    \begin{enumerate}[(1)]
    \item Significant disagreements occur among the early-type dwarf galaxies (i.e., dE, dE(N), dS0, and dS0(N)) for each catalog. Some early-type dwarf galaxies in the EVCC are classified as galaxies with late-type (i.e., BCD and Im) or unknown type morphologies in the VCC (see upper left part of Fig.~\ref{ComparVCC}). All of these galaxies exhibit elliptical shapes with overall smooth and regular appearances (see Figure~\ref{IrrtodE}). Although at a first glance galaxies like EVCC\,403, \SC{EVCC\,730}, EVCC\,836, EVCC\,851, and EVCC\,1004 seem to have a central star\SC{-}forming region, using an alternating color scheme to display the $r$-band image reveals smooth, elliptical isophotes right to the center. Other galaxies like EVCC\,464, EVCC\,806, and EVCC\,832 in Fig.~\ref{IrrtodE} seem to host many small, compact star\SC{-}forming regions. However, these star\SC{-}forming regions turned out to be faint foreground stars and background galaxies mostly with red colors judging from visual inspection of the color images.

The opposite situation is also seen : some early-type dwarf galaxies in the VCC have been classified as irregulars (see lower right part of Fig.~\ref{ComparVCC}) in the EVCC. In our classification, while some of these galaxies present elliptical shapes in outer part, they show large irregular structures with blue colors at their centers (e.g, EVCC\,187, EVCC\,246, EVCC\,301, EVCC\,316, EVCC\,374, EVCC\,400, EVCC\,1143, \SC{and EVCC\,2227} in Figure~\ref{dEtoIrr}). Besides, some other galaxies show significantly \SC{off-center} blue cores (e.g., EVCC\,159, EVCC\,245, EVCC\,653, and EVCC\,960 in Fig.~\ref{dEtoIrr}). 
    \item Combining properties like angular size and surface brightness distribution led to the reclassification of some early-type galaxies (E and S0) in the VCC to early-type dwarf galaxies (dE and dS0) in the EVCC\footnote{
    \RP{While our morphological classification of E/S0/dE/dS0 was made by visual inspection of images, based on the peak surface brightness values of galaxies returned from \SC{SExtractor}, we found that the peak surface brightness limit for discrimination between E/S0 and dE/dS0 is $\mu_{r}$ $\sim$19\,mag\,arcsec$^{-2}$ . In the R(50) vs. peak surface brightness diagram, \Suk{the} size of E/S0 galaxies decreases with peak surface brightness in \Suk{a} wide range of  0.5 $<$ log(R50) $<$ 2.0, whereas dE/dS0 \Suk{galaxies have almost a constant size in a relatively narrow range of} 0.5 $<$ log(R50) $<$ 1.5.}
    }
     (see Figure~\ref{GToD}). While the main criterion for distinguishing dE/dS0 from E/S0 has been the faint surface brightness of the former, this distinction becomes increasingly difficult at faint (M$_{r}$ $>$ -18) magnitudes.  At these magnitudes, changes from E/S0 to dE/dS0 are therefore expected between two surveys with different instruments and especially different bands, namely the blue-sensitive VCC and the multi-band EVCC. We could not see that these galaxies have a particularly high surface brightness or are particularly compact in our classification.
    
    \item The dwarf lenticular galaxies (dS0 and dS0(N)) in the EVCC have large scatter in the VCC morphological classification: about \Suk{55$\%$ (44 of 80)} of dwarf lenticular galaxies in the EVCC were classified as  different types in the VCC. A significant fraction of these galaxies is classified as E and dE types in the VCC, however, they show prominent irregularities at the center and/or asymmetric features like bar or lens in SDSS images (see Figure~\ref{LenVCC}).
    \item Some \Suk{(20 galaxies)} dEs in the VCC have been re-classified as dE(N) in the EVCC. As shown in Figure~\ref{dEN} (top panels), from SDSS images of dEs in the VCC, clear nuclei are visible. It is thought that a fraction of dEs was miss-classified in the VCC because either the resolution of the photographic plate material or the seeing was insufficient to detect a central point source. Furthermore, three dS0s in the VCC also have been assigned to the dS0(N)s in the EVCC (see middle panels of Fig.~\ref{dEN}). On the other hand, some \Suk{(19 galaxies)} dE(N)s in the VCC do not show a nuclei at their centers from SDSS images and they have been re-classified as normal dEs in the EVCC (see bottom panels of Fig.~\ref{dEN}).
    \end{enumerate}

    \subsection{Projected Spatial Distribution of Galaxies}

        \begin{figure*}
        \epsscale{0.8}
        \plotone{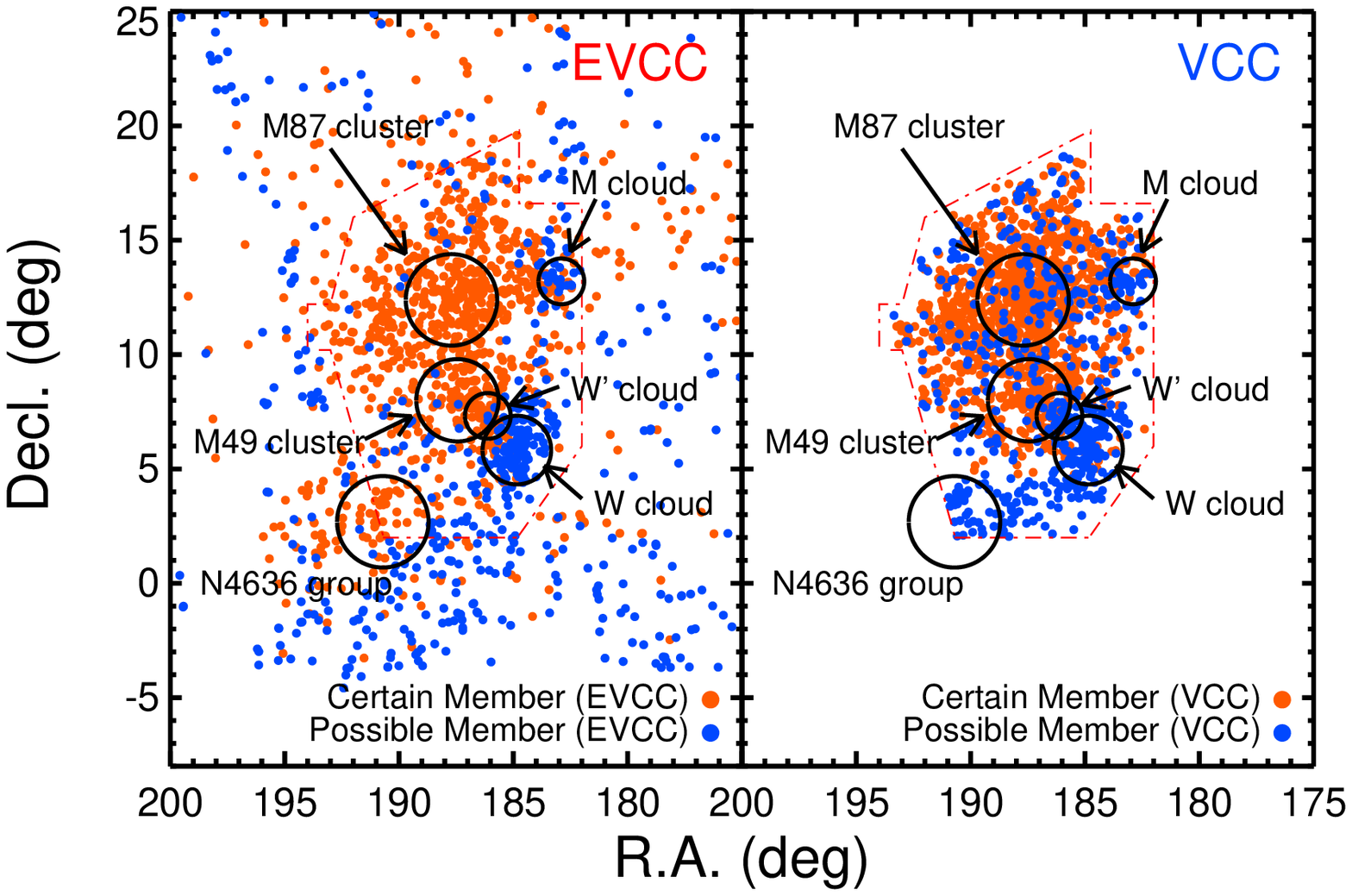}
        \centering 
        (a)
        \plotone{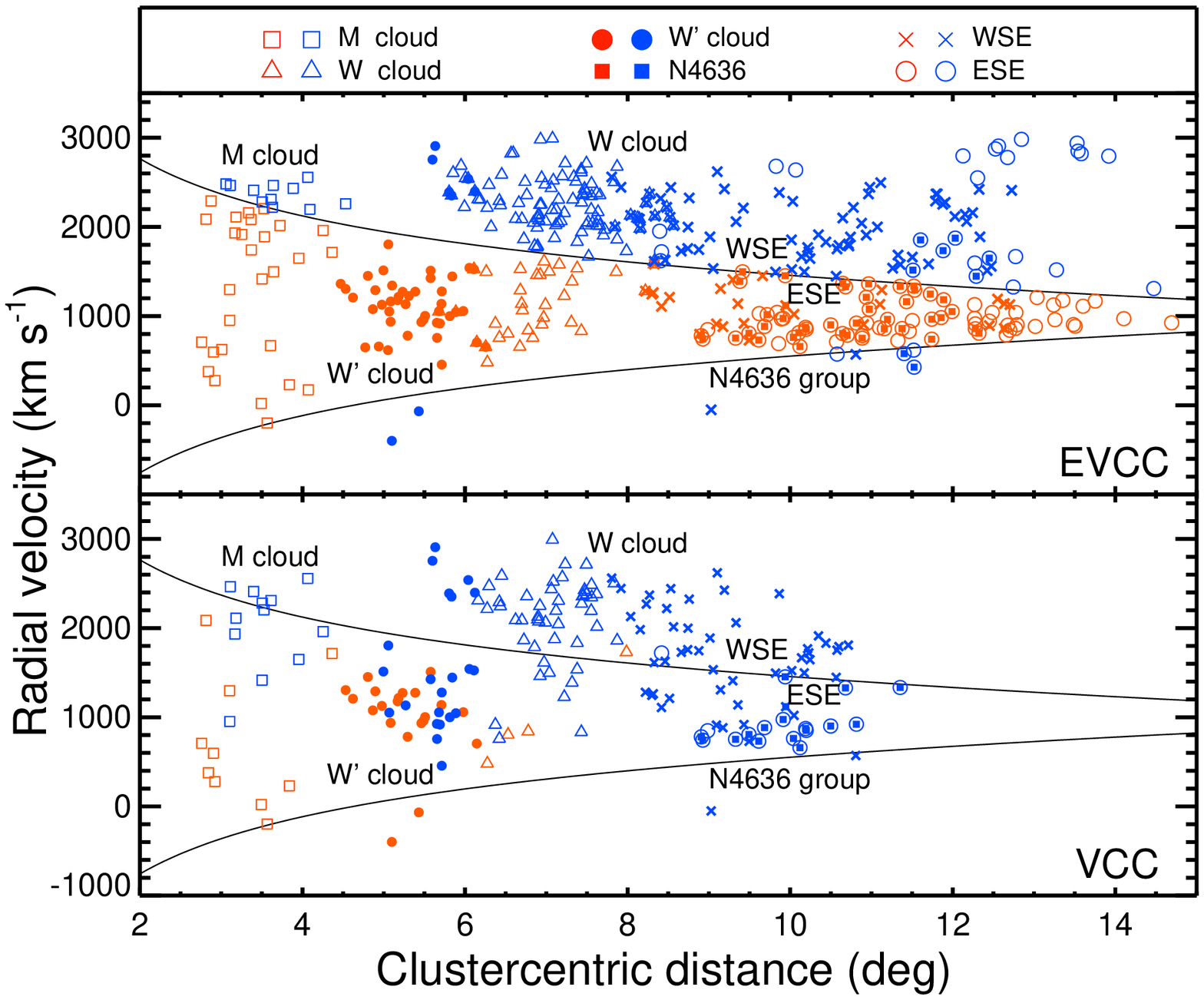} 
        (b)
        \caption{}
        \end{figure*}
        \addtocounter{figure}{-1}
        \begin{figure*}
        \caption{
        (Top) Projected spatial distribution of galaxies in the EVCC (left) and VCC (right). Red and blue filled circles are certain and possible member galaxies, respectively. The membership information of the right panel is from the VCC. The red dot dashed contour outlines the survey boundary of the VCC. The subgroups (M87, M49, M, W, and W$'$) are presented by large circles as defined by \citet{Bin87} and the NGC 4636 group is defined by this study. Note that the \Suk{size of the large circle does not represent the physical extend of each subgroup.} (Bottom) Same as Figure~\ref{Sel_member} but for galaxies in the EVCC and VCC which are associated with well-known subgroups in the Virgo cluster. Galaxies included in different subgroups are denoted as different symbols; open squares for M cloud, filled circles for W$'$ cloud, open triangles for W, open circles for eastern part of the southern extension (ESE), crosses for western part of the southern extension (WSE), and filled squares for NGC 4636 group. The ESE and WSE are defined 
        as 189 deg $<$ R.A. $<$ 197 deg , 0 deg $<$ Decl. $<$ 5 deg 
        and 184 deg $<$ R.A. $<$ 189 deg , 0 deg $<$ Decl. $<$ 5 deg, respectively. 
In each panel, red and blue color symbols are for certain and possible member galaxies defined from each catalog, respectively. 
        }

        \label{Sel_member_field}
        \end{figure*}

        \begin{figure*}
        \epsscale{0.73}
        \begin{center}
        \plotone{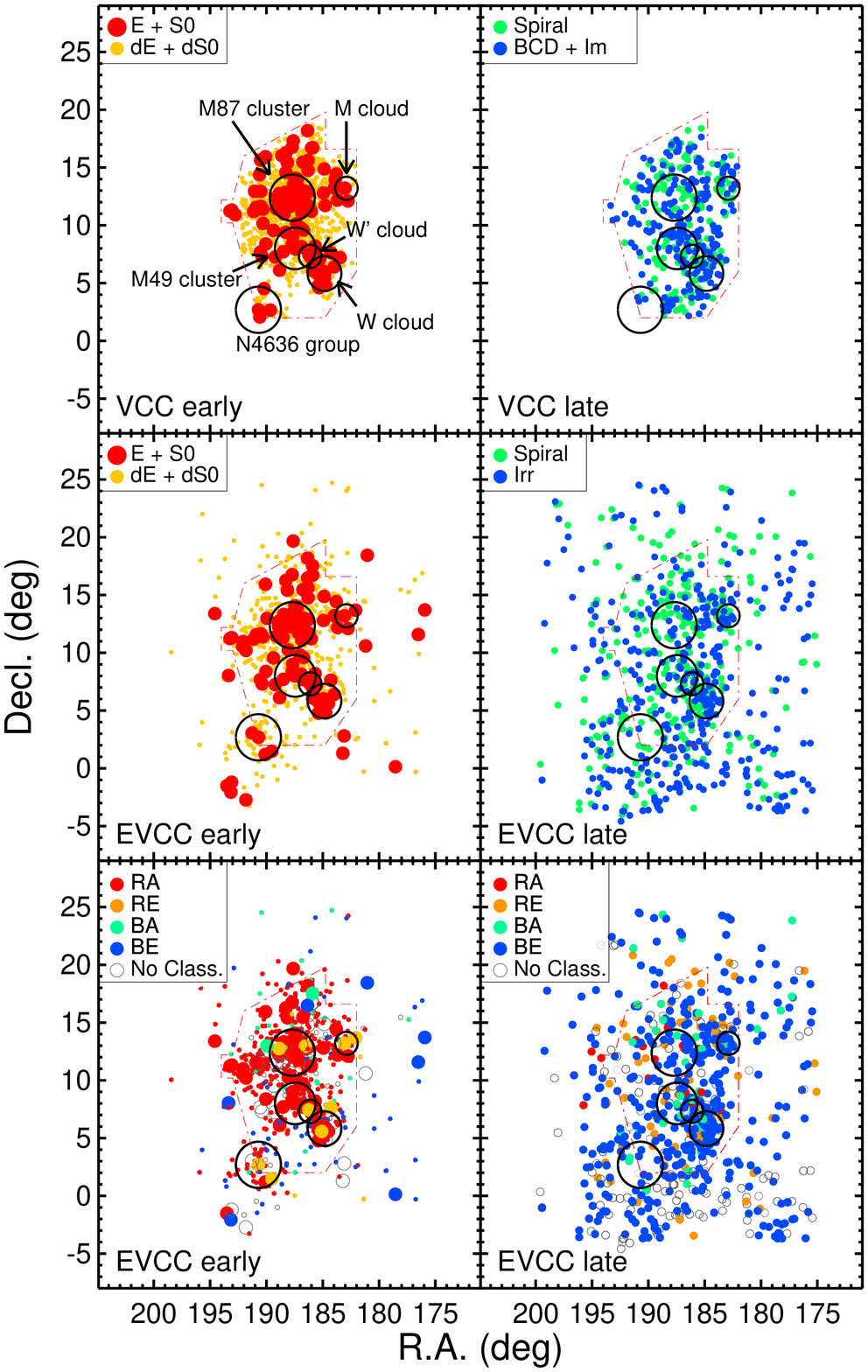}
        \caption{
        \RP{Comparison of the projected spatial distributions of galaxies between the VCC (top panels) and EVCC (middle and bottom panels) according to the galaxy morphology. Left and right panels are for early (E/S0 and dE/dS0) and late-type  (spirals and  irregulars) galaxies, respectively. In \Suk{the} bottom panels, filled circles with different colors represent different secondary morphologies in the EVCC, and open circles are galaxies with no available secondary classification information which are only compiled from the NED and have no information of SED.}
        }
        \label{psd_EVCC_p}
        \end{center}
        \end{figure*}    
		
    Fig.~\ref{Sel_member_field}a (left panel) shows the distribution of galaxies on the sky classified as certain members (red circles) and possible members (blue circles) in the EVCC. For comparison, we also present the distribution of VCC galaxies (see right panel of Fig.~\ref{Sel_member_field}a). Member galaxies in the EVCC are widely distributed beyond the region of the VCC. \Suk{Basically, the distributions of certain and possible member galaxies in the EVCC} appear to be similar to those in the VCC within the area of the VCC. As previous studies suggested \citep{San85_2,Bin87,Bin93}, subgroups (W and M) of the Virgo cluster are dominated by possible members, and this feature is also seen in the EVCC. On the other hand, the majority of EVCC galaxies in the W$'$ group are \Suk{classified} as certain members in contrast to the case of the VCC. All of galaxies in the southern extension area (Decl. $<$ 5\,deg) are defined as possible members in the VCC \citep[see also Figure 1 and 2 of][]{San85_2}. However, in the case of the EVCC, a large fraction of certain member galaxies is mixed with possible members in this region. Furthermore, a significant fraction of certain member galaxies in the EVCC is also found in the eastern part of the southern extension of the Virgo cluster, which overlaps with previously known galaxy groups \citep[e.g., NGC 4636,][]{Bro06}.

 Most galaxies in the W and M clouds are defined as possible members in the VCC since they are at their mean velocity of $\sim$2000\,km\,s$^{-1}$. They are thought to lie at about twice the distance of the Virgo cluster \citep{Bin93}. These clouds might be linked to each other \citep{Fta84,Yoo12} and are falling into the Virgo cluster from the backside \citep{Yas97}. However, \citet{Gav99} argue that the W cloud is a rather isolated structure following the Hubble flow,\RP{ \Suk{whereas the M cloud appears to fall into the cluster.}}  In the EVCC, the W cloud (\RP{open triangles} in \SC{the} upper panel of Fig.~\ref{Sel_member_field}b) appears to be a localized, independent structure in which a large fraction of galaxies resides outside of the caustic curves of the infall model lines. On the other hand, the majority of galaxies in the M cloud turn out to be certain members which are inside of the infall models (open squares in \SC{the} \RP{upper panel} of Fig.~\ref{Sel_member_field}b). In this case, we suggest that the M cloud consists of a galaxy population infalling into the Virgo cluster and spatially projected galaxies associated with main body of the cluster.

\RP{\citet{Bin87} proposed that the W$'$ cloud might be in between the W cloud and the cluster B.} \Suk{It has been further suggested that the} W$'$ cloud is connected to the background W cloud as a filament and both are falling into the cluster B \citep{Bin93}. However, \citet{Mei07} proposed that the W$'$ could be a more localized structure behind the Virgo cluster based on the distance estimation of five galaxies ($\sim$22.9\,$\pm$\,0.3\,Mpc in the mean). \RP{In the EVCC, most galaxies associated with the W$'$ cloud are well inside of the infall model lines and have \Suk{a} mean velocity ($\sim$1140\,km\,s$^{-1}$) \Suk{comparable to the systemic Virgo cluster velocity} (filled circles in \SC{the} upper panel of Fig.~\ref{Sel_member_field}b). Consequently, considering its estimated distance \citep{Mei07} combined with our velocity distribution, \Suk{we suggest that galaxies in the W$'$ cloud are falling into the} Virgo cluster from behind.} 

    To the south of the Virgo cluster (i.e., Decl. $<$ 5\,deg), the southern extension defines an extended structure which seems to be a filament in the Local Supercluster \citep{deV73,Tul82,Hof95}. The southern extension is also suggested as an infalling structure to the Virgo cluster \citep[e.g.,][]{Bin87}. Furthermore, the western part of this structure might be linked to the W and M clouds \citep{Yoo12}. In the VCC, galaxies in the southern extension were classified either as possible member or as background object. However, in the EVCC, some galaxies in the southern extension are classified as certain members. Note that such certain members are preferentially located on the eastern side of the southern extension (hereafter ESE), while the western side (hereafter WSE) is dominated by possible members (see also left panel of Fig.~\ref{Sel_member_field}a). While the galaxies associated with the ESE are far from the Virgo center, most of them are well inside of the narrow range of the infall model lines. The ESE is mostly associated with NGC\,4636 group which exhibits intragroup X-ray emission and is dynamically mature \citep{Bro06}. Previous studies suggest that NGC\,4636 group is also infalling into the Vigo cluster \citep{Tul84,Nol93}.
  
    In Figure~\ref{psd_EVCC_p}, we present the projected spatial distribution of galaxies in the EVCC (middle and bottom panels) in comparison with those in the VCC (top panels) according to the morphology of galaxies (left and right panels for early- and late-type galaxies, respectively). 
    The spatial distribution of E/S0 galaxies in the EVCC appear to be similar to that of the VCC. \RP{A Kolmogorov-Smirnov (KS) test gives the probability of 99\% that the spatial distributions of E/S0 \Suk{galaxies in the two catalogs are drawn} from the same distribution. \Suk{This is expected as the morphological classifications of most early-type galaxies between the EVCC
and VCC agree} (see Sec. 6.1).}
Early-type galaxies are strongly concentrated around M87 and the centers of subgroups: \RP{about 60\% of early-type and 30\% of late-type galaxies are located within \Suk{an} angular distance of 2 deg (corresponding to 575\,kpc) from M87 and subgroups.} Furthermore, they appear to be surrounded by early-type dwarf galaxies (dEs and dS0s): \RP{about 80\% of dE/dS0 galaxies reside within \Suk{an} angular distance of 1\,deg (corresponding to 288\,kpc) from E/S0 galaxies.}
    One interesting feature is that the EVCC reveals a \Suk{population} of E/S0 galaxies around and south of the NGC 4636 group, a region that was not covered by the VCC. 
     \RP{It is clear that early-type and late-type galaxies in the EVCC are dominated by RA/RE and BA/BE type galaxies, respectively (see bottom panels of Fig.~\ref{psd_EVCC_p}): a KS test gives that \SC{the} spatial distributions of RA/RE and BA/BE type galaxies are almost similar to those of early-type and late-type galaxies at the 97\% and 95\% confidence level, respectively. The early-type galaxies with red spectra (RA and RE types) \Suk{seem to concentrate around the main body} and subgroups of the Virgo cluster, whereas those with BA and BE types are preferentially located in their outskirts: about 85\% of early-type galaxies residing within the main body and subgroups are those of RA and RE types.}

    Late-type galaxies (spirals and irregulars) are relatively more scattered and randomly distributed compared to the early-type galaxies. Furthermore, the region outside of the VCC survey area, which is covered only by the EVCC, are dominated by late-type galaxies. Most of the subgroups (W, W$'$, and M clouds) contain many late-type galaxies whereas the NGC\,4636 group lacks these galaxies. 

    \subsection{Luminosity Function}

        \begin{figure}
        \epsscale{1}
        \begin{center}
        \plotone{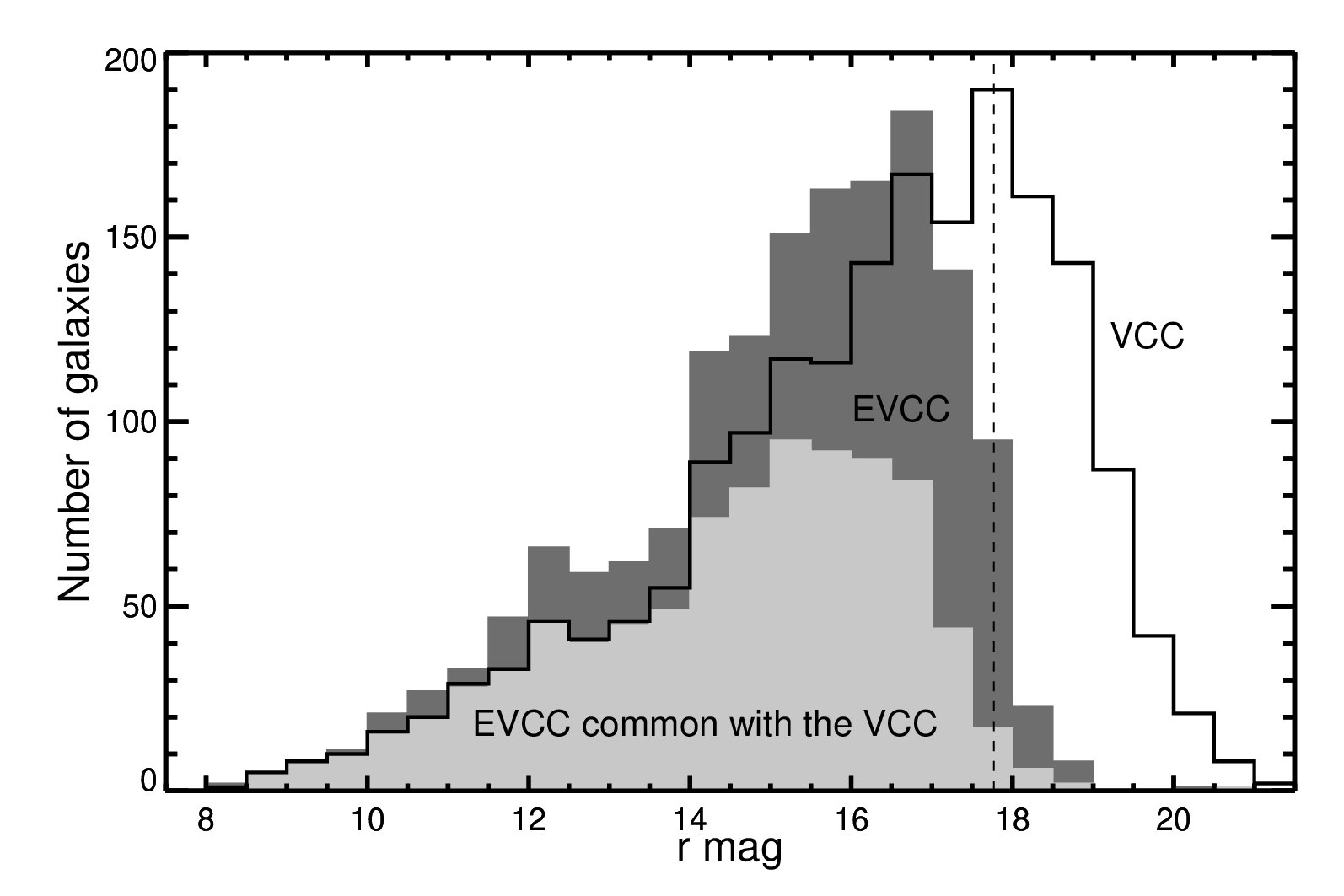}
        \caption{ 
        Galaxy luminosity function in the EVCC (dark grey histogram) in comparison with 
        that of the VCC (open histogram). Open histogram is for all certain and possible member galaxies in the VCC. Light grey histogram is for the EVCC galaxies that are in common with the VCC. The vertical dashed line is the limiting magnitude of the SDSS spectroscopic survey (i.e., $r =$ 17.7).
        }
        \label{TLF}
        \end{center}
        \end{figure}
		
        \begin{figure}
        \epsscale{1}
        \begin{center}
        \plotone{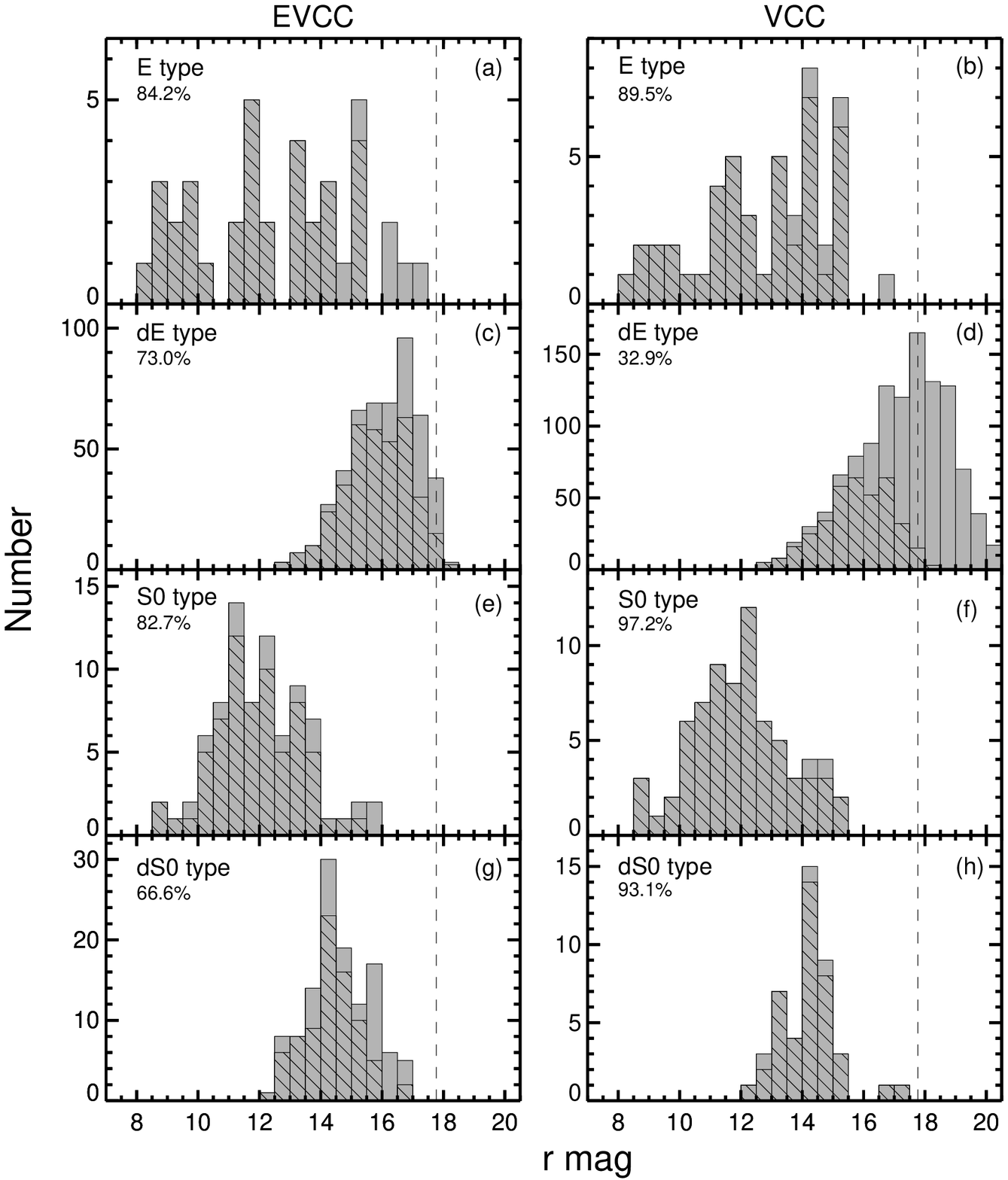}
        \caption{ 
        Galaxy luminosity function of early-type galaxies (E, dE, S0, and dS0) in the EVCC (left panels) and VCC (right panels). Certain and possible member galaxies are considered. In each panel, \RP{the} hatched histogram is for galaxies included both in the EVCC and  the VCC, and its fraction is also indicated below the galaxy morphological type.
        } 
        \label{LFearly}
        \end{center}
        \end{figure}
		
        \begin{figure}
        \epsscale{1}
        \begin{center}
        \plotone{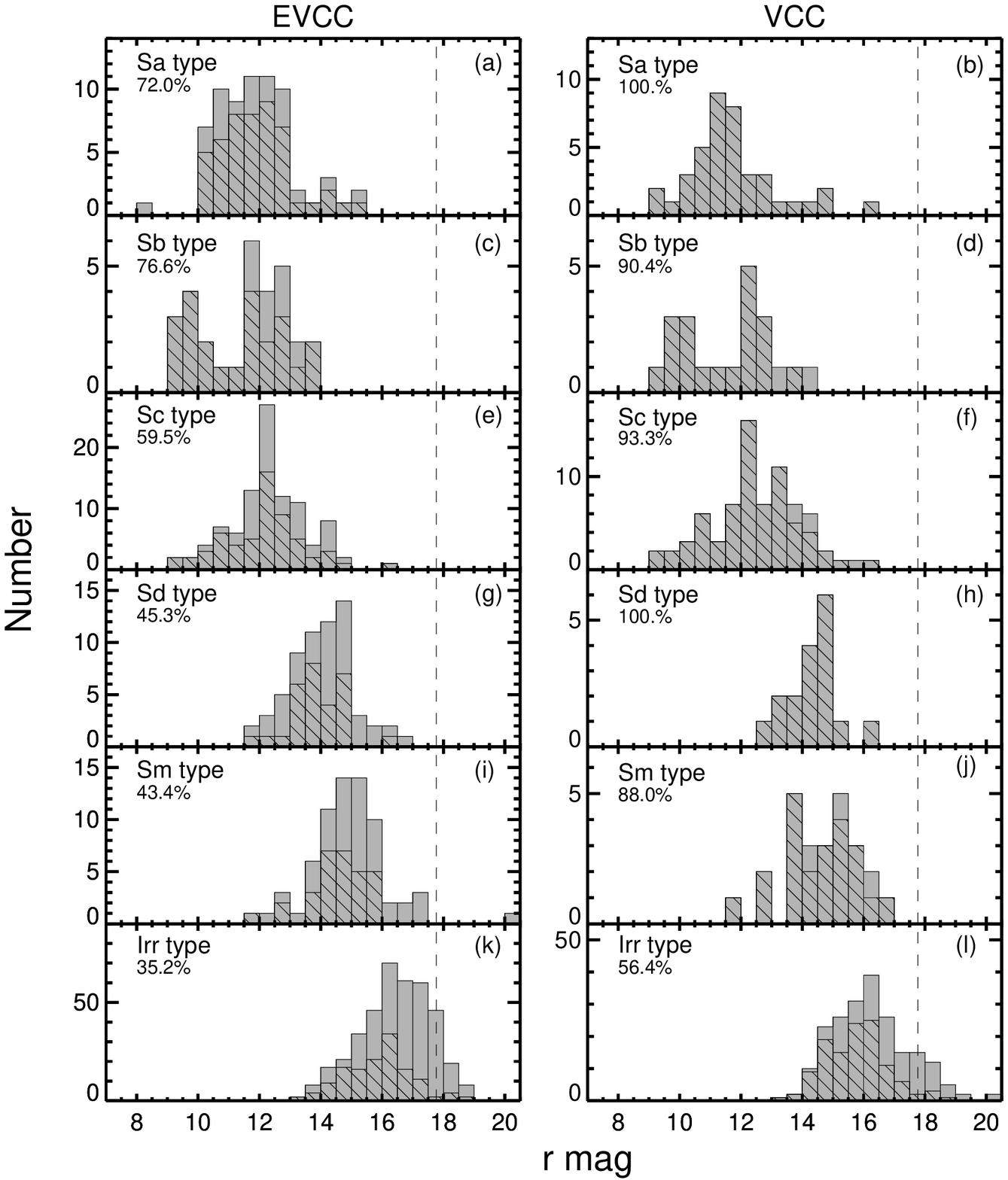}
        \caption{
        Same as Fig.~\ref{LFearly} but for late-type galaxies (Sa, Sb, Sc, Sd, Sm, and Irr) 
        in the EVCC (left panels) and VCC (right panels).
        }
        \label{LFlate}
        \end{center}
        \end{figure}

    Figure~\ref{TLF} shows the r-band galaxy luminosity function in the EVCC in comparison with that of the VCC. The luminosity function of EVCC galaxies does not include galaxies fainter than $r\approx19$, which is mainly due to the limiting magnitude of SDSS spectroscopic observations (i.e.,  $r =$ 17.7, vertical dashed line). Furthermore, the apparent incompleteness of the luminosity function is also seen at brighter magnitude (r $\sim$16.5) than the nominal spectroscopic limit, which is consistent with the decrease of the S/P ratio in Fig.~\ref{fraction}. As expected from the type-specific galaxy luminosity functions \citep{Bin88,Fer88,Jer97_1}, these faint samples in the VCC are found to be dE and Irr galaxies (see also Figure~\ref{LFearly} and Figure~\ref{LFlate}). The mean magnitude of galaxies that are exclusive to the EVCC (dark grey histogram) is $\sim$1.6\,mag fainter than that of galaxies that are in common with the VCC (light grey histogram). This indicates that the EVCC galaxies located outside of the VCC region have systematically lower luminosities. This is a consequence of the morphology-density relation where the low density environments lack the luminous early-type galaxies found in the inner regions of galaxy clusters. 

    In Fig.~\ref{LFearly} and Fig.~\ref{LFlate}, we present galaxy luminosity function in the EVCC (left panels) and VCC (right panels) for various galaxy morphological types. In each panel, the hatched histogram is for galaxies included both in the EVCC and VCC. First of all, for a given galaxy morphological type, the shape of the luminosity function of the EVCC appear to be comparable to that of the VCC in the sense of similar magnitude range and mean magnitude of the distribution. It is shown in Fig.~\ref{LFlate} that the mean magnitude of the distribution changes with galaxy type. Especially, the mean magnitude of late-type galaxies systematically varies in the sense that the Irr and Sm galaxies are fainter than spiral galaxies and the Sd galaxies are fainter than the Sa, Sb, and Sc galaxies  \citep{San85_2,Bin88,Fer88,Jer97_1}. 

    In the case of galaxies in the VCC (right panels of Fig.~\ref{LFearly} and Fig.~\ref{LFlate}), most galaxies are recovered by the EVCC. However, about \Suk{67$\%$} and \Suk{42$\%$} of dE and Irr galaxies in the VCC, respectively, are not included in the EVCC. A considerable fraction of these galaxies is fainter than the limiting magnitude of the SDSS spectroscopic survey. On the other hand, some dE and Irr galaxies brighter than this magnitude limit are not also found in the  EVCC. While these galaxies are certain and possible members in the VCC, they are excluded in the EVCC owing to their larger radial velocities (i.e., $>$ 3000\,km\,s$^{-1}$). In the case of galaxies in the EVCC (left panels of Fig.~\ref{LFearly} and Fig.~\ref{LFlate}), at all magnitudes, some EVCC galaxies are not found in the VCC. This is resulted from that most of them are located outside of the VCC region. This feature is more distinct for the Sm and Irr type galaxies, since these galaxies are preferentially found in the outskirts of clusters.

	\subsection{Astrometry and Membership of VCC Galaxies}
	
        \begin{figure}
        \epsscale{0.87}
        \begin{center}
        \epsscale{0.9}
        \plotone{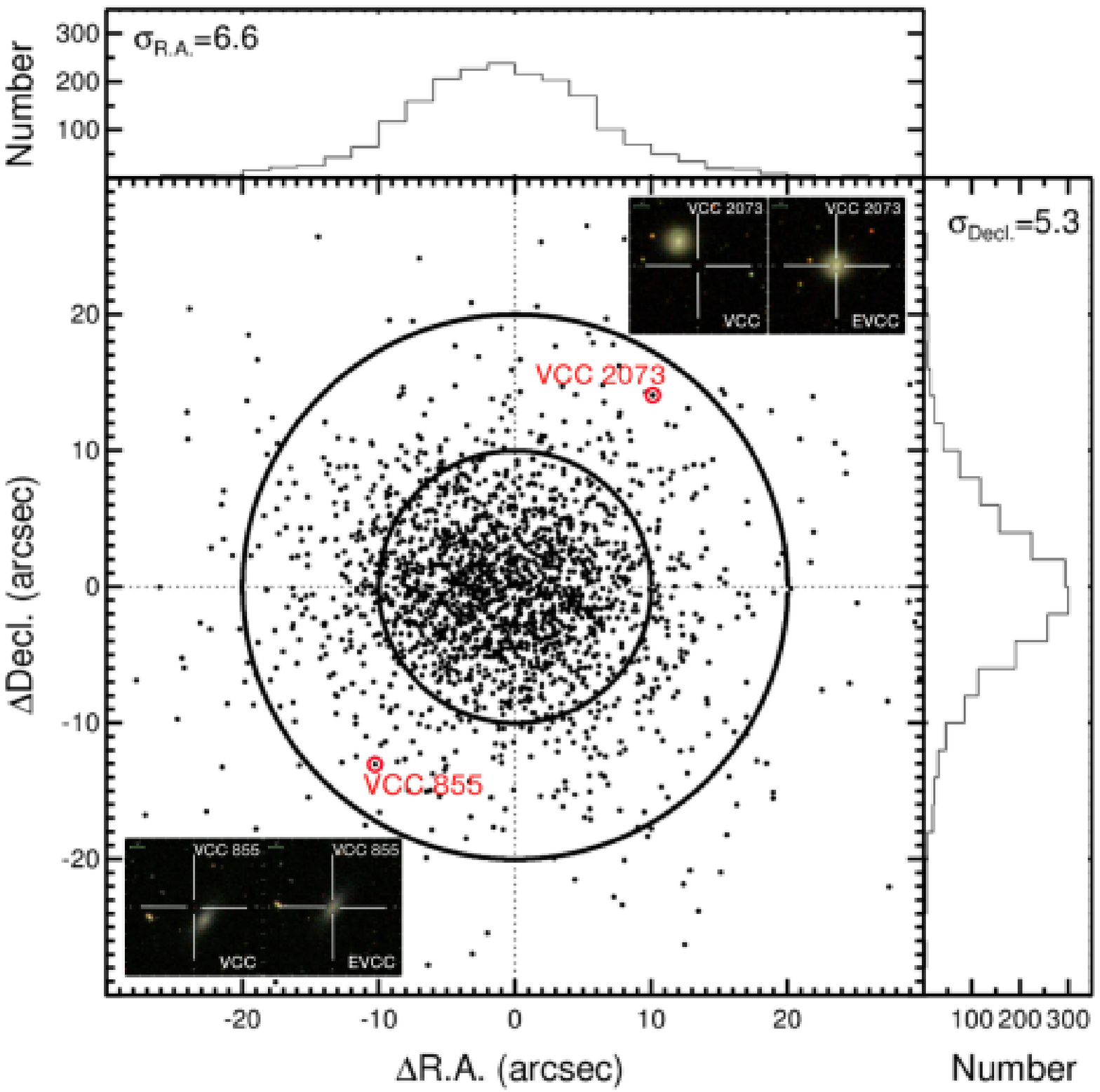}
        \caption{
        Residual of astrometry between the EVCC and VCC. Two circles of 10 and 20 arcsec radii are superposed to the plot as references. The histograms are the distributions of the residuals in R.A. (top panel) and Decl. (right panel). The standard deviation of the residual distribution (in arcsec) is presented in each panel. Inset images are examples of galaxies (VCC 855 and VCC 2073) with wrong coordinates in the VCC.
        }
        \label{astrometry}
        \end{center}
        \end{figure}
		
        \begin{figure}[htp]
        \centering
        \epsscale{1}
        \plotone{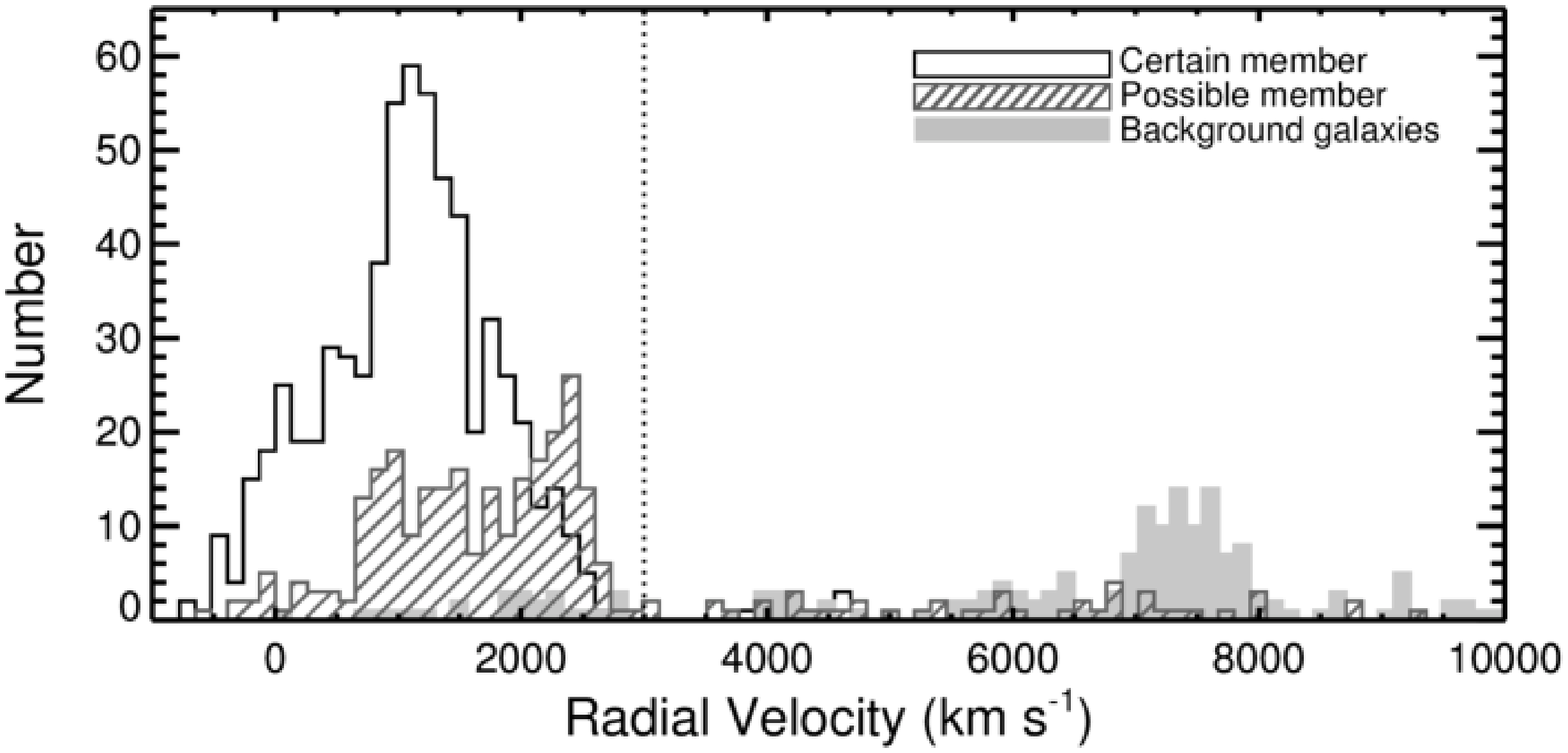}
        \caption{
        Radial velocity distribution of galaxies in the VCC with available spectra. Solid, hatched, and filled histograms are certain member, possible member, and background galaxies, respectively, defined by the VCC. Vertical dotted line is for the limit of 3000 km\,s$^{-1}$.
        }
        \label{member}
        \end{figure}
		
    By matching coordinates of all galaxies presented in the VCC with those from SDSS images, we have found that coordinates of large numbers of galaxies in the VCC significantly deviate from our coordinates and are discrepant from true centers of galaxies. Moreover, a considerable fraction of galaxies in the NED also shows incorrect coordinates. For VCC galaxies with incorrect coordinates, by taking into account of their size of D25, magnitude, and morphology, we identify real object from the visual inspection of SDSS images.  As a result, about 28$\%$ of galaxies in the VCC (586 of 2096) turned out to have different coordinates from the EVCC with offset larger than 10\,arcsec (see Figure~\ref{astrometry} for residuals of coordinates between the EVCC and VCC). In our EVCC (see Table 2 and Table 3), coordinates of all galaxies in the VCC and NED  have been superseded by those returned from our own photometry. 

        \begin{table*}[hpt]
        \caption{VCC Galaxies with Redshift Data}
        \begin{tabular}{  c  c  c  c  c }
        \hline
        &  VCC Certain Member  &  VCC Possible Member  &  VCC Background  & Total  \\  \hline
        $cz$ $<$ 3000 km\,s$^{-1}$  &  631  &  257  &  25  &  913 \\  
        $cz$ $>$ 3000 km\,s$^{-1}$  &  84  &  96  &  221  &  401  \\  
        Total  &  715  &  353  &  246  &  1314  \\  \hline
        \end{tabular}\label{pm}
        \end{table*}	

    The cluster membership of galaxies in the VCC was solely based on morphology, confirmed by the velocity data available for some galaxies \citep{Bin85}. In Table 4, we compare the membership of VCC galaxies with radial velocity information in the EVCC (see also Figure~\ref{member}). We divide galaxies into those with larger or lower velocities than the velocity limit of 3000\,km\,s$^{-1}$. First of all, only about \Suk{12\% (84/715)} of certain member galaxies in the VCC are found to have larger radial velocities than this velocity limit. In the case of VCC possible members, about \Suk{27\% (96/353)} of galaxies turn out to have radial velocities larger than 3000\,km\,s$^{-1}$. On the other hand, about \Suk{10\% (25/246)} of background objects in the VCC are within this limit of radial velocity. \citet{Bin93} estimated that, based on the velocity information available for 144 galaxies, the success rate for the morphology-based cluster membership of the VCC is 95\%. However, when we carefully examined the details of the Table 1 of \citet{Bin93}, we found that the change of membership status is slightly high compared to the result presented in Table 4: $\sim$13\% (6 of 45) from certain member to background, $\sim$33\% (31 of 94) from possible member to background, and $\sim$20\% (1 of 5) from background to certain member. 

        \begin{figure}
        \epsscale{0.8}
        \begin{center}
        \plotone{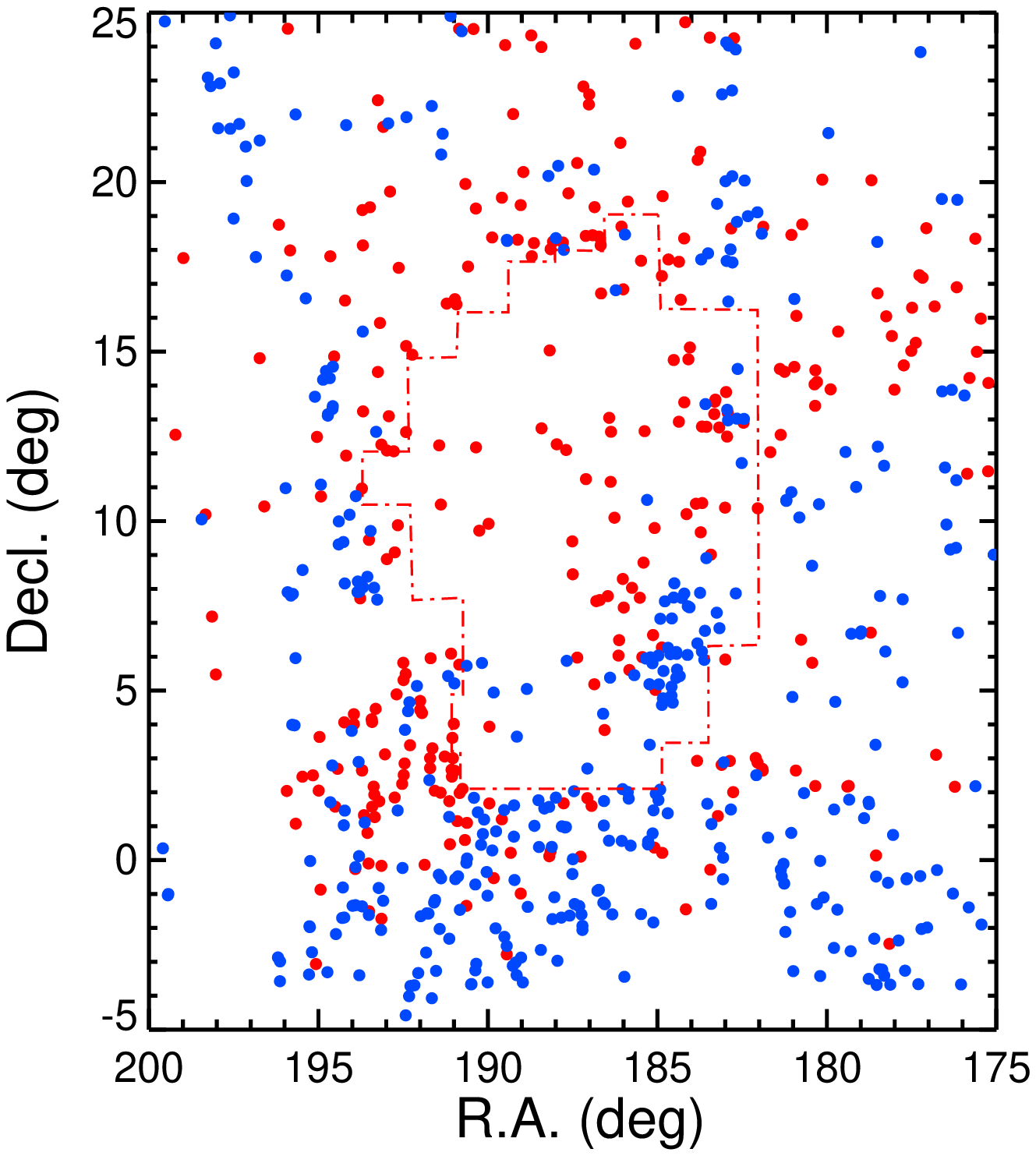}
        \caption{
        Projected spatial distribution of 676 galaxies included only in the EVCC. Red and blue filled circles are certain and possible member galaxies, respectively. The red contour marks the VCC footprint.
        }
        \label{onlyEVCCpsd}
        \end{center}
        \end{figure}

        \begin{figure}
        \epsscale{0.8}
        \begin{center}
        \plotone{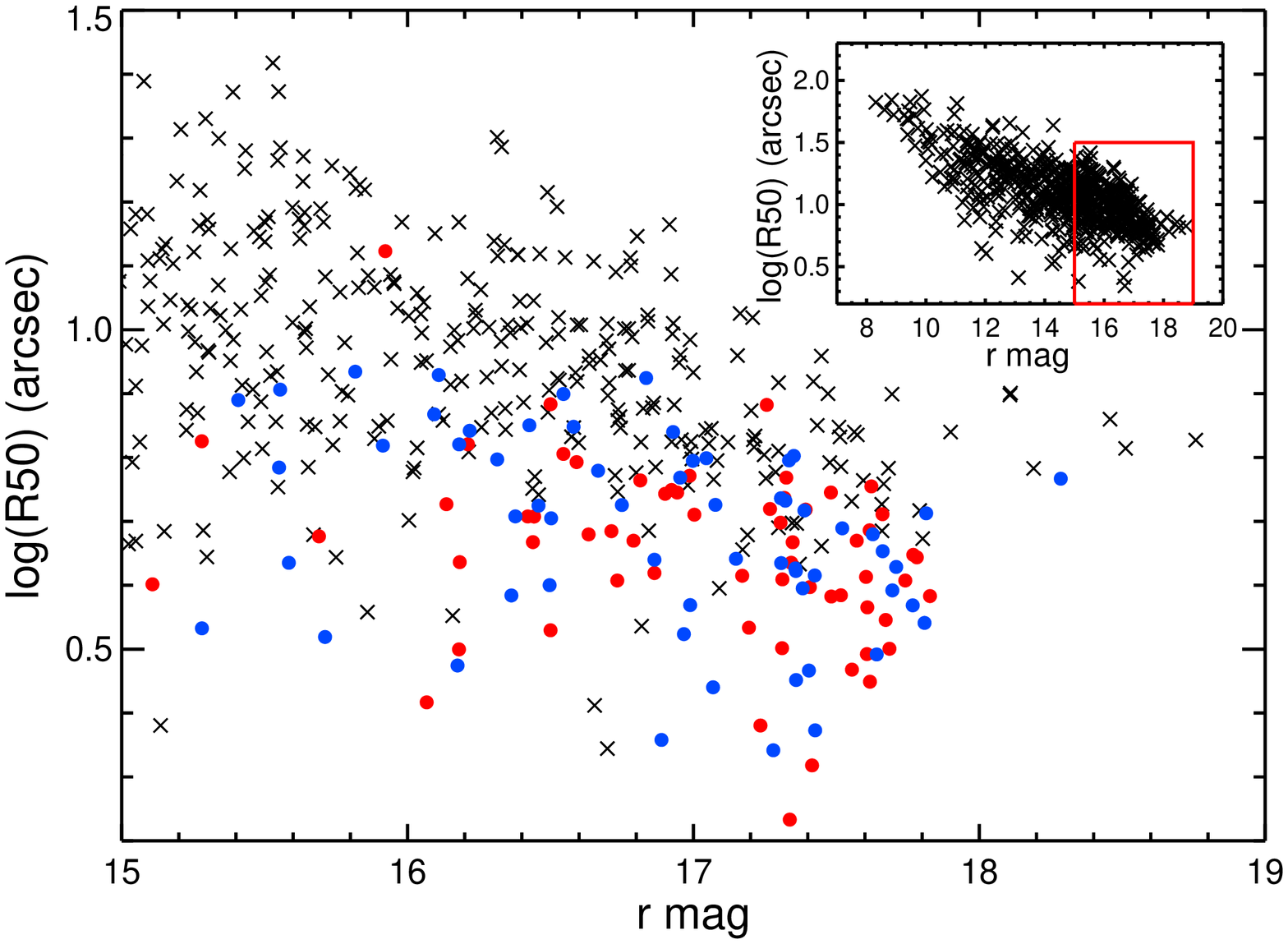}
        \caption{
        The $r$-band half-light radius (R50) versus magnitude for faint EVCC galaxies located in the region of the VCC. Red and blue filled circles are certain and possible member galaxies included only in the EVCC, respectively. Crosses are VCC certain and possible member galaxies.  The inset shows all EVCC galaxies of entire magnitude range located in the VCC region.
        }
        \label{onlyEVCCr50}
        \end{center}
        \end{figure}
		
    In Fig.~\ref{onlyEVCCpsd}, we present the projected spatial distribution of 676 galaxies that are exclusive to the EVCC. While most of these galaxies are distributed outside of the VCC region, about 19$\%$ of galaxies are still located inside of the VCC region. About 51$\%$ and 49$\%$ of the latter are certain and possible member galaxies, respectively. It is interesting to note that these new galaxies have typically magnitudes fainter than $r >$ 15 and show morphologies of Irr and dE/dS0. Furthermore, these galaxies show systematically smaller effective radii at a fixed magnitude than VCC galaxies (see Figure ~\ref{onlyEVCCr50}). This indicates that these galaxies are more compact with  centrally concentrated light distribution when compared to the majority of galaxies in the VCC. We therefore suspect that the comparatively poor angular resolution of the VCC photographic plates prevented a morphology-based membership assignment of these galaxies in the VCC. In addition, given that these more compact galaxies bare a resemblance to background galaxies could be the reason why they were rejected as cluster members in the VCC study. 
\\

\section{Summary and Conclusions}
\Suk{We have} presented a new catalog of galaxies in the Virgo cluster and wider area, the Extended Virgo Cluster Catalog (EVCC), based on the SDSS DR7 and NED data. The EVCC covers a rectangular region (\Suk{60.1\,Mpc$^{2}$} assuming a distance of the Virgo cluster of 16.5\,Mpc) which is \Suk{5.2 times} larger than the footprint of the classical VCC. This paper focuses on a broad overview of the construction of the EVCC including selection of sample galaxies as well as their morphological classification and measurements of photometric parameters. The main results of this paper are summarized as follows:

\begin{enumerate}[1.]
    \item We focus on the sample galaxies with radial velocities that are available in the SDSS and NED spectroscopic data, taking advantage of radial velocity as a suitable information for cluster membership for galaxies in a galaxy cluster. We thus selected all galaxies in the EVCC region with radial velocities less than 3000\,km\,s$^{-1}$ as a mean to separate member galaxies from background. This process secured a total of 1589 galaxies of which 676 galaxies are not part of the VCC. The member galaxies are divided into two subsamples using a spherical symmetric infall model in a plot of radial velocity versus clustercentric distance of galaxies. Certain and possible Virgo cluster members are defined as those located inside and outside of the model lines, respectively. 

    \item We introduced two independent, complementary morphological classification schemes using the SDSS imaging and spectroscopic data: the primary morphology and secondary morphology. The primary morphology is based on the extended Hubble morphological classification scheme used in the VCC. The secondary morphological classification is using the SED shape and the presence of H$\alpha$ emission/absorption lines. The EVCC galaxies were subdivided into 21 classes in the primary and  four classes in the secondary morphology.

    \item The primary and secondary morphologies show an expected general trend in the sense that early-type galaxies exhibit red spectra and the majority of late-type galaxies show blue spectra, although some scatters in the main trend between two morphologies are shown. This correlation implies that the secondary morphology, which is based on the spectrum of the central region of galaxy, also reflects global characteristics of the galaxies. Furthermore, we found that within our classification scheme, fiber location in the SDSS does not significantly affect the secondary morphology at a given primary morphology.
    
    \item The final EVCC catalog includes fundamental information of galaxies such as membership, morphology, and photometric parameters. In addition, we also computed photometric parameters for \Suk{1183} galaxies in the VCC which are not in common with the EVCC. The photometry of all galaxies was performed using our own SExtractor pipeline in order to remedy deficiencies in the SDSS photometry caused by deblending and sky subtraction. 
    
    \item The primary morphology in the EVCC is generally consistent with the VCC classification in which a tight trend along the diagonal between two catalogs is found. The relationship has some outliers at a given morphological type where the morphologies of galaxies in our EVCC disagree with those in the VCC. However, it is worth to note that, in most cases, the morphologies of these galaxies have been updated in the EVCC thanks to the SDSS images with higher resolution and S/N ratio compared to the photographic plate images of the VCC.
    
    \item Member galaxies in the EVCC are widely distributed beyond the region of the VCC. Basically, the certain and possible member galaxies in the EVCC show \SC{the} projected spatial distributions similar to counterparts in the VCC within the area of the VCC. The W subgroup of the Virgo cluster is dominated by possible members in the EVCC. On the other hand, the majority of EVCC galaxies in the W$'$ subgroup turned out to be certain members in contrast to the case of the VCC. A large fraction of certain member galaxies in the EVCC is found in the southern extension, while all of VCC galaxies in this region are defined as possible members. Especially, a significant fraction of certain member galaxies in the EVCC is distributed toward the eastern part of the southern extension including the previously known galaxy group of NGC 4636.

    \item The overall spatial distribution of EVCC galaxies according to the morphology is similar to that of the VCC. The early-type galaxies are strongly concentrated around M87 and the centers of subgroups, whereas late-type galaxies are more evenly distributed. Outside of the VCC area the galaxy populations are dominated by late-type galaxies. It is worth to note that the EVCC reveals many early-type galaxies in the eastern part of the southern extension (i.e., around and below NGC\,4636 group), a region that was not covered by the VCC.

    \item While the EVCC contains galaxies down to about $r =$ 19, some fainter galaxies in the morphology-based VCC are absent in the EVCC. This is mainly due to the limiting magnitude of SDSS spectroscopic observations and these faint samples are mostly dE and Irr galaxies. For a given galaxy morphological type except dE and Irr, the shape of the luminosity function of the EVCC is similar to that of the VCC.
    
    \item We have found that coordinates of large numbers of galaxies in the VCC significantly are deviated from our coordinates and are discrepant from true centers of galaxies. About 28$\%$ of galaxies in the VCC turned out to have different coordinates from the EVCC with offset larger than 10\,arcsec. 

    \item Based on the limit of the radial velocity of 3000\,km\,s$^{-1}$ as a membership criterion in the EVCC, about \Suk{12$\%$} and \Suk{27$\%$} for certain and possible member galaxies, respectively, in the VCC have larger velocities than 3000\,km\,s$^{-1}$. On the other hand, about \Suk{10$\%$} of background galaxies in the VCC are within this radial velocity limit. About 19$\%$ of 676 newly covered galaxies in the EVCC are located inside of the VCC region, while they were not selected as member galaxies in the VCC. These galaxies have faint luminosities and are smaller in size compared to the galaxies in the VCC. They show morphologies of Irr and dE/dS0.
    \end{enumerate}
    
Our EVCC not only updates and extends the VCC, but also will be complementary to other available or ongoing surveys of the Virgo cluster at various wavelengths (see \citealt{Fer12} for the review), and serves as basis for forthcoming scientific studies of the Virgo cluster. Taking advantages of the EVCC catalog in terms of homogeneous data for unprecedented wide-field region of the Virgo cluster as well as available radial velocity and spectroscopic information of all member galaxies, our forthcoming papers will explore a number of scientific issues, including the following.

    \begin{enumerate}[(1)]
    \item Kinematic structure of the Virgo cluster and its dependence on various environmental parameters
    \item Variation of luminosity functions of galaxies in various regions  
    \item Chemical properties of extensive emission line galaxies as a function of various cluster environments
    \item Ultraviolet properties of galaxies and correlations with environments
    \item Surface brightness profiles and structural parameters of galaxies 
    \item Identification of potential "feeding channels" of the Virgo cluster by using dwarf galaxy properties to relate known cluster substructures to regions beyond the virial radius
    \item 3D-structure of the Virgo environment from surface brightness fluctuations for early-type galaxies
    \end{enumerate}

\acknowledgments
We are grateful to the anonymous referee for helpful comments and suggestions that improved the clarity and quality of this paper. This research was supported by Basic Science Research Program through the National Research Foundation of Korea (NRF) funded by the Ministry of Education, Science, and Technology (NRF-2012R1A1B4003097). Support for this work was also provided by the NRF of Korea to the Center for Galaxy Evolution Research (No. 2010-0027910). 
S.K. acknowledges support from the National Junior Research Fellowship of NRF (No. 2011-0012618).
T.L. was supported within the framework of the Excellence Initiative by the German Research Foundation
(DFG) through the Heidelberg Graduate School of Fundamental Physics (grant number GSC 129/1).

\bibliography{ms}

\end{document}